\documentclass[vecphys]{svmult}

\usepackage{makeidx}         
\usepackage{graphicx}        
\usepackage{multicol}        
\usepackage[bottom]{footmisc}

\begin{document}

\title*{Superconducting Polarons and Bipolarons}
\author{A. S. Alexandrov}
\institute{Department of Physics, Loughborough University, Loughborough LE11 3TU, United Kingdom \\
\texttt{a.s.alexandrov@lboro.ac.uk}}
%
%
\maketitle

\begin{abstract}
The seminal work by Bardeen, Cooper and Schrieffer (BCS)\index{BCS
theory}
 extended further by Eliashberg  to the intermediate coupling regime solved one of the major scientific
problems of Condensed Matter Physics in the last century. The BCS
theory provides qualitative and in many cases quantitative
descriptions of low-temperature superconducting metals and their
alloys,
 and some novel high-temperature superconductors  like
magnesium diboride. The theory has been extended by us to the
strong-coupling regime where carriers are small lattice polarons and
bipolarons. Here I review the multi-polaron strong-coupling theory
of superconductivity. \index{superconductivity!strong-coupling}
Attractive electron correlations, prerequisite to any
superconductivity, are caused by an almost unretarded
electron-phonon (e-ph) interaction sufficient to overcome the direct
Coulomb repulsion in this regime. Low energy physics is that of
small polarons and bipolarons, which are real-space electron (hole)
pairs dressed by phonons. They are itinerant quasiparticles existing
in the Bloch states at temperatures below the characteristic phonon
frequency. Since there is almost no retardation (i.e. no
Tolmachev-Morel-Anderson logarithm) reducing the Coulomb repulsion,
e-ph interactions should be relatively strong to overcome the direct
Coulomb repulsion, so carriers \emph{must} be polaronic to form
pairs in novel superconductors. I identify  the long-range
 Fr\"{o}hlich
electron-phonon interaction \index{Fr\"ohlich!interaction} as the
most essential for pairing in superconducting cuprates. A number of
key observations have been predicted or explained  with polarons and
bipolarons including unusual isotope effects and upper critical
fields, normal state (pseudo)gaps and kinetic properties, normal
state diamagnetism, and giant proximity effects. These and many
other observations provide strong evidence for a novel state of
electronic matter in layered cuprates, which is a charged
Bose-liquid of small mobile bipolarons.

\end{abstract}

\section{ \label{sec:one}
Introduction } While a single polaron problem has been actively
researched for a long time (for reviews see ref.
\cite{alemot,dev,rash,mih,sal,sto,app,fir,bot} and  the present
volume.), multi-polaron physics has gained particular attention in
the last two decades. For weak electron-phonon coupling, $\lambda <
1$, and the adiabatic limit, $\omega /E_F \ll 1$), Migdal theory
\index{Migdal theory}describes electron dynamics in the normal
Fermi-liquid state \cite{migdal1958a}, and BCS-Eliashberg theory in
the superconducting state \cite{bcs,eli} (here and further I use
$\hbar=c=k_B=1$). While the electron-phonon (e-ph) interaction is
weak  Migdal's theorem
 is perfectly applied. The theorem proves that the
contribution of diagrams with "crossing" phonon lines (so called
"vertex" corrections) is small if the parameter $\lambda \omega/
E_{F}$ is small, where $\lambda$ is the dimensionless (BCS) e-ph
coupling constant, $\omega$ is the characteristic phonon frequency,
and $E_{F}$ is the Fermi energy. Neglecting the vertex corrections,
Migdal \cite{migdal1958a} calculated the renormalized electron mass
as $m^*=m_0(1+\lambda)$ (near the Fermi level), where $m_0$ is the
band mass in the absence of e-ph interaction, and Eliashberg
\cite{eli} extended Migdal's theory to describe the BCS
superconducting state at intermediate values of $\lambda < 1$. Later
on many researchers applied Migdal-Eliashberg theory
\index{Migdal-Eliashberg theory} with $\lambda$ even larger than 1
(see, for example, Ref.~\cite{sca}, and references therein). With
increasing strength of interaction and increasing phonon frequency,
$\omega$, finite bandwidth \cite{alemaz,dog} and vertex corrections
\cite{hag0}
 become increasingly important. But unexpectedly for many
 researchers who applied the non-crossing
approximation  even at $\lambda >1$ we have found that
 the Migdal-BCS-Eliashberg theory (with or without vertex corrections)
breaks down entirely at $\lambda \sim 1$ for any value of the
adiabatic ratio $\omega/E_F$ since the bandwidth is narrowed and the
Fermi energy, $E_F$  is renormalised down exponentially so the
effective parameter $\lambda\omega/E_F$ becomes large \cite{ale0}.

The electron-phonon coupling constant $\lambda $
\index{electron-phonon!coupling constant} is about the ratio of the
electron-phonon interaction energy $E_{p}$ to the half bandwidth $%
D\approx N(E_{F})^{-1}$, where $N(E)$ is the density of electron
states in a rigid lattice. One  expects
 \cite{ale0} that when the
coupling is strong, $\lambda
>1$, all electrons in the bare Bloch band are "dressed" by phonons
since their kinetic energy ($<D$) is small compared with the
potential energy due to the local lattice deformation, $E_{p}$,
caused by electrons. In this strong coupling regime the canonical
Lang-Firsov transformation \cite{lang1962a} \index{Lang-Firsov
transformation} can be used to determine the properties of the
system. Under certain conditions
\cite{alemulty,aleFermi,aleeur,alebook}, the multi-polaron system is
metallic but with polaronic carriers rather than bare electrons.
This regime is beyond Migdal-Eliashberg theory, where the effective
mass approximation is used and the electron bandwidth is infinite.
In particular, the small polaron regime cannot be reached by
summation of the standard Feynman-Dyson perturbation diagrams using
a translation-invariant Green's function $G({\bf r},{\bf
r'},\tau)=G({\bf r}-{\bf r'},\tau)$ with the Fourier transform
$G({\bf k}, \Omega)$ prior to solving the Dyson equations on a
discrete lattice.  This assumption excludes the possibility of local
violation of the translational symmetry \cite{land} due to the
lattice deformation in any order of the Feynman-Dyson perturbation
theory similar to the absence of the anomalous (Bogoliubov) averages
in any order of perturbation theory \cite{migdal1958a}. To enable
electrons to relax into the lowest polaronic band, one has to
introduce an infinitesimal translation-noninvariant potential, which
should be set zero only in the final solution obtained by the
summation of Feynman diagrams for the Fourier transform $G({\bf
k},{\bf k'},\Omega)$ of $G({\bf r},{\bf r'},\tau)$ rather than for
$G({\bf k}, \Omega)$ \cite{alemaz}. As in the case of the
off-diagonal superconducting order parameter, the off-diagonal terms
of the Green function, in particular the Umklapp terms with ${\bf
k'}={\bf k+G}$, drive the system into a small polaron ground state
at sufficiently large coupling.  Setting the
translation-noninvariant potential  to zero in the solution of the
equations of motion restores the translation symmetry but in a
polaron band rather than in the bare electron band, which turns out
to be an excited state.

Alternatively, one can work with momentum eigenstates throughout the
whole coupling region, but taking into account the finite electron
bandwidth from the very beginning. In recent years many numerical,
and analytical  studies have confirmed the conclusion \cite{ale0},
that the Migdal-Eliashberg theory breaks down at $\lambda \geq
 1$ (see, for example
Refs.~\cite{kab,kab2,feh,mar,tak,feh2,tak2,rom,lam,zey,wag,aub,trugman,Korn2}
and contributions to this book). With increasing phonon frequency
the range of validity of the $1/\lambda$ polaron expansion extends
to smaller values of $\lambda$ \cite{ale2}.  As a result, the region
of applicability of the Migdal-Eliashberg approach (even with vertex
corrections) shrinks to smaller values of the coupling, $\lambda <
1$, with increasing $\omega$.  Strong correlations between carriers
reduce this region further (see \cite{feh2}).

 Carriers in the fascinating advanced
materials are strongly coupled with high-frequency optical phonons,
making small polarons and  non-adiabatic effects relevant for
high-temperature superconductivity, colossal magnetoresistance
phenomenon, and molecular electronic devices (see Part III). Indeed
the characteristic phonon energies $0.05-0.2$ eV in cuprates,
manganites and in many organic materials are of the same order as
generally accepted values of the hopping integrals $0.1 - 0.3$ eV.

As reviewed in this book lattice polarons are different from
ordinary electrons in many aspects, but  perhaps one of the most
remarkable difference is found in their superconducting properties.
By extending the BCS theory towards the strong interaction between
electrons and ion vibrations, a charged Bose gas (CBG)
\index{charged Bose gas} of tightly bound  small bipolarons was
predicted by us \cite{aleran} with a further prediction  that high
critical temperature $T_c$ is found in the crossover region of the
e-ph interaction strength from the BCS-like \emph{polaronic}  to
\emph{bipolaronic} superconductivity \cite{ale0}. This contribution
describes what happens to the conventional BCS theory when the
electron-phonon coupling becomes strong. The author's particular
view of  cuprates is also presented.

\section{Electron-phonon and Coulomb interactions in Wannier representation}

For doped semiconductors and metals with a strong electron-phonon
 interaction it is convenient to transform the Bloch states $|\bf k\rangle$ to
the site (Wannier) states $|\bf m\rangle$ using the canonical linear
transformation of the electron operators,\index{Wannier
representation}
\begin{equation}
c_{i}={\frac{1}{\sqrt{N}}}\sum_{{\bf k}}e^{i{\bf k\cdot m}}c_{{\bf
k}s},
\end{equation}
where $i=({\bf m},s)$ includes both site ${\bf m}$ and spin $s$
quantum numbers, and $N$ is the number of sites in a crystal. In the
site representation the electron kinetic energy takes the following
form
\begin{equation}
H_{e}=\sum_{i,j}\left[ T({\bf m-m^{\prime }})\delta _{ss^{\prime
}}-\mu \delta _{ij}\right] c_{i}^{\dagger }c_{j},
\end{equation}
where
\[
T({\bf m})={\frac{1}{{N}}}\sum_{{\bf k}}E_{{\bf k}}e^{i{\bf k\cdot
m}}
\]
is the bare hopping integral in the rigid lattice,  $\mu$ is the chemical potential,  $j=(%
{\bf n},s^{\prime })$, and $E_{{\bf k}}$ is the bare Bloch band
dispersion.

The electron-phonon and Coulomb interactions acquire simple forms in
the Wannier representation, if their matrix elements in the momentum
representation $\gamma ({\bf q,}\nu )$ and $V_{c}({\bf q})$ depend
only on the momentum transfer ${\bf q}$,
\begin{equation}
H_{e-ph}=\sum_{{\bf q},\nu ,i}\omega _{{\bf q}\nu }\hat{n}_{i}\left[ u_{i}(%
{\bf q,}\nu )d_{{\bf q}\nu }+H.c.\right] ,
\end{equation}
and
\begin{equation}
H_{e-e}={\frac{1}{{2}}}\sum_{i\neq j}V_{c}({\bf
m-n})\hat{n}_{i}\hat{n}_{j}.
\end{equation}
Here
\begin{equation}
u_{i}({\bf q,}\nu )={\frac{1}{\sqrt{2N}}}\gamma ({\bf q,}\nu )e^{i{\bf %
q\cdot m}}
\end{equation}
and
\begin{equation}
V_{c}({\bf m})={\frac{1}{{N}}}\sum_{{\bf q}}V_{c}({\bf q})e^{i{\bf q\cdot m}%
},
\end{equation}
are the matrix elements of e-ph and Coulomb interactions,
respectively, in the Wannier representation for electrons,  $\hat{n}%
_{i}=c_{i}^{\dagger }c_{i}$ is the electron density operator, and
$d_{{\bf q}\nu }$  annihilates the $\nu$-branch phonon with the wave
vector $\bf q$ and frequency $\omega _{{\bf q}\nu }$. Taking the
interaction matrix elements depending only on the momentum transfer
one neglects terms in the electron-phonon and Coulomb interactions,
which are proportional to the overlap integrals of the Wannier
orbitals on different sites. This approximation is justified for
narrow band materials with the bandwidth $2D$  less than the
characteristic value of the crystal field. As a result,  the generic
Hamiltonian takes the following form in the Wannier representation,
\begin{eqnarray}
H &=&\sum_{i,j}\left[ T({\bf m-m^{\prime }})\delta _{ss^{\prime
}}-\mu
\delta _{ij}\right] c_{i}^{\dagger }c_{j}+\sum_{{\bf q},\nu ,i}\omega _{{\bf %
q\nu }}\hat{n}_{i}\left[ u_{i}({\bf q,}\nu )d_{{\bf q}\nu
}+H.c.\right]
\nonumber \\
&&+{\frac{1}{{2}}}\sum_{i\neq j}V_{c}({\bf m-n})\hat{n}_{i}\hat{n}_{j}+\sum_{%
{\bf q}}\omega _{{\bf q}\nu }(d_{{\bf q}\nu }^{\dagger }d_{{\bf
q}\nu }+1/2).
\end{eqnarray}
Here we  confine our discussions to
 a single electron band  and  the e-ph matrix element  depending only on the momentum transfer
${\bf q}$. This approximation allows for qualitative and  in many
cases quantitative descriptions of essential polaronic effects in
advanced materials. There are might be degenerate atomic orbitals in
solids coupled to local molecular-type Jahn-Teller distortions,
where one has to consider multi-band electron energy structures (see
\cite{kornil}).

The quantitative calculation of the matrix element in the whole
region of  momenta
 has to be performed from  pseodopotentials
\cite{shue,max,bar}. On the other hand  one can parameterize the
e-ph interaction rather than to compute it from first principles in
many physically important cases \cite{mahan}. There are three most
important interactions in doped semiconductors, which are polar
coupling to optical phonons (i.e the Fr\"ohlich e-ph interaction),
deformation potential coupling to acoustical phonons, and the local
(Holstein) e-ph interaction \index{Holstein!interaction} with
molecular type vibrations in complex lattices. While  the matrix
element is ill defined in metals,  it is well defined in doped
semiconductors, which have their parent dielectric compounds,
together with bare phonons $\omega _{{\bf q}\nu }$ and the electron
band structure $E_{ {\bf k}}$ Here the effect of carriers on the
crystal field and on the dynamic matrix is small while the carrier
density is much less than the atomic one (for phonon self-energies
and  frequency renormalizations in polaronic systems see Ref.
\cite{alemulty,kabanov}).  Hence one can use the band structure and
the crystal field of parent insulators to calculate the matrix
element in doped semiconductors.

The e-ph matrix element $\gamma ({\bf q})$ has different
$q$-dependence for different phonon branches. In the long wavelength
limit ($q\ll \pi /a$, $a$ is the lattice constant), $\gamma ({\bf
q}) \propto q^n$, where $n=-1,0$ and $n=-1/2$ for polar optical,
molecular ($\omega _{{\bf q}}=\omega_0)$) and acoustic ($(\omega
_{{\bf q}}\propto q)$) phonons, respectively.  Not only $q$
dependence is known but also the absolute values of $\gamma ({\bf
q})$ are well parameterized in this limit. For example in polar
semiconductors the interaction of two slow electrons at some
distance $r$ is found as (see below)
\begin{equation}
v(r)=V_c(r)-{1\over{N}}\sum_{\bf q} |\gamma({\bf q})|^2 \omega_{\bf
q} e^{i{\bf q \cdot r}}.
\end{equation}
The Coulomb repulsion in a rigid lattice is
$V_c(r)=e^2/\epsilon_\infty r$, and the second term represents the
difference between the Coulomb repulsion screened with  the core
electrons  and the repulsion screened with both core electrons and
ions.  Hence the matrix element of the
 Fr\"ohlich interaction depends only on the dielectric constants and the optical phonon frequency $\omega_0$
 as
 \begin{equation}
|\gamma({\bf q})|^2={4\pi e^2\over{\kappa \omega_{0}}},
\end{equation}
where $\kappa= (\epsilon_\infty^{-1}-\epsilon_0^{-1})^{-1}$.

One can transform the e-ph interaction further using the
site-representation also for phonons. The site representation of
phonons is particularly convenient for the interaction with
dispersionless local modes,
whose $\omega _{{\bf q\nu }}=\omega _{\nu }$ and the polarization vectors ${\bf e}_{{\bf q}\nu }={\bf %
e}_{\nu }$ are ${\bf q}$ independent. Introducing the phonon
site-operators
\begin{equation}
d_{{\bf n\nu }}={\frac{1}{\sqrt{N}}}\sum_{{\bf k}}e^{i{\bf q\cdot n}}d_{{\bf %
q\nu }}
\end{equation}
one transforms the deformation energy and the e-ph interaction as
\cite{alebook}
\begin{equation}
H_{ph} =\sum_{{\bf n,\nu }}\omega _{\nu }(d_{{\bf n}\nu }^{\dagger }d_{%
{\bf n}\nu }+1/2),
\end{equation}
and
\begin{equation}
H_{e-ph}=\sum_{{\bf n,m,}\nu }\omega _{\nu }g_{\nu }({\bf m-n})({\bf e}_{%
{\bf \nu }}\cdot {\bf e}_{{\bf m-n}})\hat{n}_{{\bf m}s}(d_{{\bf
n}\nu }^{\dagger }+d_{{\bf n}\nu }),
\end{equation}
respectively.

Here $ g_{\nu }({\bf m})$
is a dimensionless {\em force} acting between the electron on site{\bf \ }$%
{\bf m}$ and the displacement of ion ${\bf n}$, and  ${\bf e}_{{\bf m-n}%
}\equiv ({\bf m-n})/|{\bf m-n}|$ is the unit vector in the direction
from the electron ${\bf m}$ to the ion ${\bf n.}$  This real space
representation  is convenient in modelling the electron-phonon
interaction in complex lattices. Atomic orbitals of an ion
adiabatically follow its motion. Therefore the electron does not
interact with the displacement of the ion, whose orbital it
occupies, that is $g_{\nu }(0)=0$.

\section{Breakdown of Migdal-Eliashberg theory in the strong-coupling regime}

Obviously a perturbative approach to the e-ph interaction fails when
$\lambda
>1.$ However one might expect that the self-consistent Migdal-Eliashberg
(ME) theory is still valid in the strong-coupling regime because it
sums the infinite set of particular non-crossing diagrams in the
electron self-energy. One of the problems with such an extension of
the ME theory is a lattice instability. The same theory applied to
phonons yields the renormalised phonon frequency \index{phonon
frequency!renormalised} $\tilde{\omega}=\omega (1-2\lambda )^{1/2}$
\cite{migdal1958a}. The frequency turns out to be zero at $\lambda
=0.5$. Because of this lattice instability Migdal \cite{migdal1958a}
and Eliashberg \cite{eli} restricted the applicability of their
approach to $\lambda <1$. However, it was  shown later that there
was no lattice instability, but only a small renormalisation of the
phonon frequencies of the order of the adiabatic ratio, $\omega /\mu
\ll 1,$ \index{adiabatic ratio} for $any$ value of $\lambda ,$ if
the adiabatic Born-Oppenheimer approach was properly applied
\cite{gei}. The conclusion was that the Fr\"{o}hlich Hamiltonian
correctly describes the electron self-energy for any value of
$\lambda ,$ but it should not be applied to further renormalise
phonons.

In fact, ME theory cannot be applied at $\lambda
>1$ for
the reason, which has nothing to do with the lattice instability. Actually the $%
1/\lambda $ muti-polaron expansion technique \cite{alemulty} shows
that the many-electron system collapses into the small polaron (or
bipolaron) regime at $\lambda \approx 1$ for any adiabatic ratio.

To illustrate the point let us compare the Migdal solution of the
simple  molecular-chain Holstein model \cite{hol} \index{Holstein!
model} with the exact solution \cite{aleeur} in the adiabatic limit,
$\omega_0 /t \rightarrow 0$ , where $t=T(a)$ is the nearest
neighbour hopping integral. The Hamiltonian of the model is
\begin{eqnarray}
H &=&-t\sum_{<ij>}c_{i}^{\dagger }c_{j}+H.c.+2(\lambda
kt)^{1/2}\sum_{i}x_{i}c_{i}^{\dagger }c_{i} \\
&&+\sum_{i}\left( -{\frac{1}{{2M}}}{\frac{\partial ^{2}}{{\partial x_{i}^{2}}%
}}+{\frac{kx_{i}^{2}}{{2}}}\right) ,  \nonumber
\end{eqnarray}
where $x_{i}$ is the normal coordinate of the molecule (site) $i$,
and $k=M\omega ^{2}$. The Migdal theorem is exact in this limit.
Hence in the framework of the
Migdal-Eliashberg theory one would expect the Fermi-liquid behaviour above $%
T_{c}$ and the BCS ground state below $T_{c}$ at any value of
$\lambda .$ In fact, the exact ground state is a self-trapped
insulator at any filling of the band, if $\lambda \geq 1.$

First we consider a two-site case (zero dimensional limit),
$i,j=1,2$ with one electron, and than generalise the result for an
infinite lattice with many electrons. The transformation
$X=(x_{1}+x_{2})$, $\xi =x_{1}-x_{2}$ allows us to eliminate the
coordinate $X$, which is coupled only with the total density
($n_{1}+n_{2}=1$). That leaves the following Hamiltonian to be
solved in the extreme adiabatic limit $M\rightarrow \infty $:
\begin{equation}
H=-t(c_{1}^{\dagger }c_{2}+c_{2}^{\dagger }c_{1})+(\lambda
kt)^{1/2}\xi (c_{1}^{\dagger }c_{1}-c_{2}^{\dagger
}c_{2})+{\frac{k\xi ^{2}}{{4}}}.
\end{equation}
The solution is
\begin{equation}
\psi =(\alpha c_{1}^{\dagger }+\beta c_{2}^{\dagger })\left|
0\right\rangle ,
\end{equation}
where
\begin{equation}
\alpha ={\frac{t}{{[t^{2}+((\lambda }kt{)^{1/2}\xi +(t^{2}+\lambda
kt\xi ^{2})^{1/2})^{2}]^{1/2}}}},
\end{equation}
\begin{equation}
\beta =-{\frac{(\lambda kt)^{1/2}\xi +(t^{2}+\lambda kt\xi ^{2})^{1/2}}{{%
[t^{2}+((\lambda kt)^{1/2}\xi +(t^{2}+\lambda kt\xi ^{2})^{1/2})^{2}]^{1/2}}}%
},
\end{equation}
and the energy is
\begin{equation}
E={\frac{k\xi ^{2}}{{4}}}-(t^{2}+\lambda kt\xi ^{2})^{1/2}.
\end{equation}
In the extreme adiabatic limit the displacement $\xi $ is classical,
so the ground state energy $E_{0}$ and the ground state displacement
$\xi _{0}$ are obtained by minimising Eq.(18) with respect to $\xi
$. If $\lambda \geq 0.5 $ we obtain
\begin{equation}
E_{0}=-t(\lambda +{\frac{1}{{4\lambda }}}),
\end{equation}
and
\begin{equation}
\xi _{0}=\left[ {\frac{t(4\lambda ^{2}-1)}{{\lambda k}}}\right]
^{1/2}.
\end{equation}
The symmetry-breaking "order" parameter is
\begin{equation}
\Delta \equiv \beta ^{2}-\alpha ^{2}={\frac{[2\lambda +(4\lambda
^{2}-1)^{1/2}]^{2}-1}{{[2\lambda +(4\lambda ^{2}-1)^{1/2}]^{2}+1}}}.
\end{equation}
If $\lambda <0.5,$ the ground state is translation invariant,
$\Delta =0, $ and $E_{0}=-t,$ $\xi =0,$ $\beta =-\alpha $. Precisely
This state is the "Migdal" solution of the Holstein model, which is
symmetric (translation invariant)  with $|\alpha |=|\beta |$. When
$\lambda <0.5$, the Migdal solution is the $only$ solution. However, when $%
\lambda >0.5$ this solution is $not$ the ground state of the system,
Fig.1. The system collapses into a localised adiabatic polaron
trapped on the "right" or on the "left-hand" site due to the finite
lattice deformation $\xi _{0}\neq 0$.
\begin{figure}[tbp]
\begin{center}
\includegraphics[angle=-90,width=0.70\textwidth]{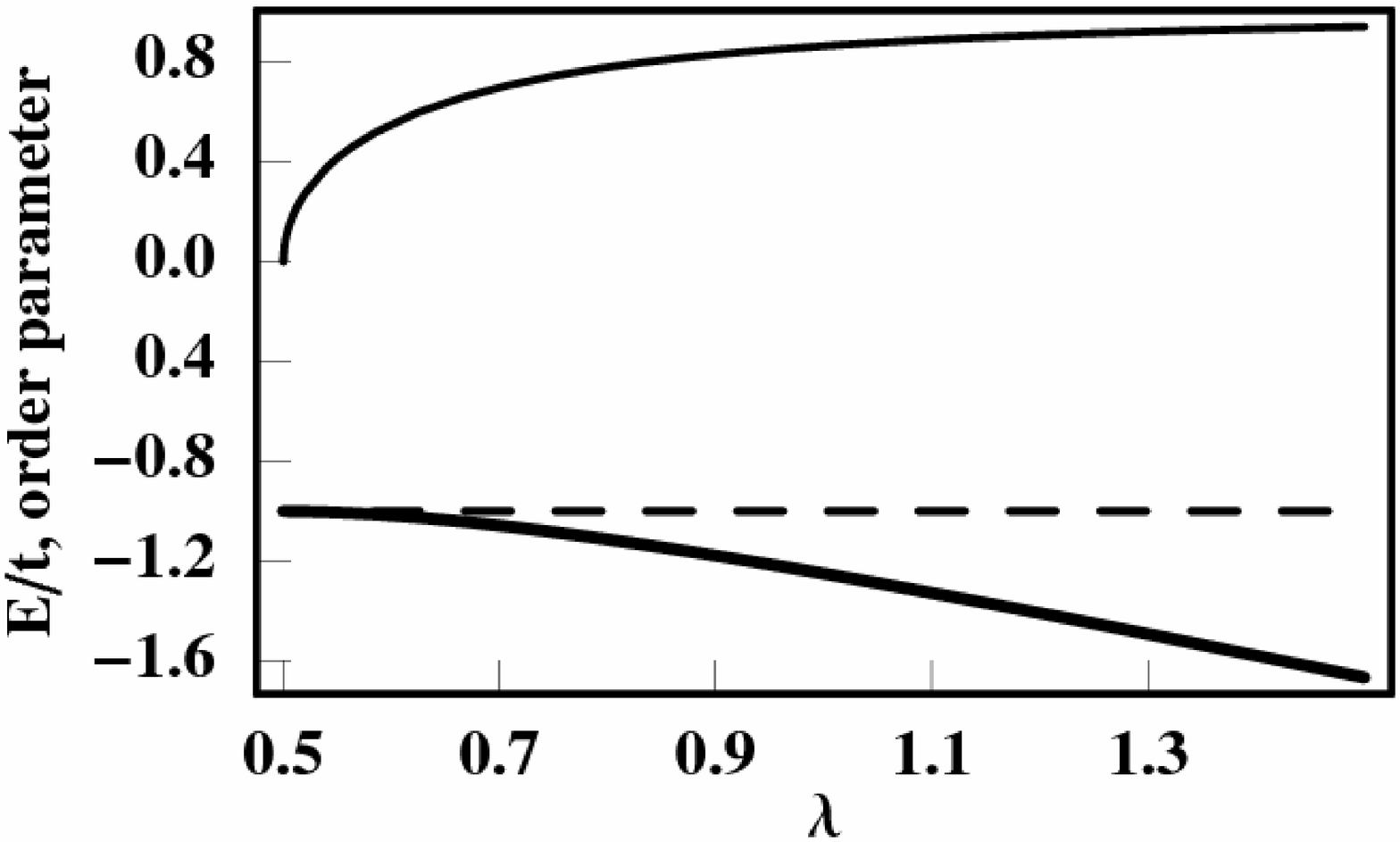}
\end{center}
\caption{The ground-state energy (in units of $t$, solid line) and
the order parameter (thin solid line) of the adiabatic Holstein
model. The Migdal solution is shown as the dashed line}
\end{figure}

The generalisation for a multi-polaron system on the infinite
lattice of any dimension is straightforward in the extreme adiabatic
regime. The adiabatic solution of the infinite one-dimensional (1D)
chain with one electron was obtained by Rashba \cite{ras} in a
continuous (i.e. effective mass) approximation, and by Holstein
\cite{hol} and Kabanov and Mashtakov \cite {kab,kabanov} for a
discrete lattice. The latter authors  studied the Holstein
two-dimensional (2D) and three-dimensional (3D) lattices in the
adiabatic limit. According to Ref. \cite{kab} the self-trapping of a
single electron occurs at $\lambda \geq 0.875$ and at $\lambda \geq
0.92$ in 2D and 3D, respectively. The radius of the self-trapped
adiabatic polaron, $r_{p}$, is readily derived from its continuous
wave function \cite{ras}
\begin{equation}
\psi (r)\sim 1/\cosh (\lambda r/a).
\end{equation}
It becomes smaller than the lattice constant, $r_{p}=a/\lambda $ for $%
\lambda \geq 1$. Hence the multi-polaron system remains in the
self-trapped \index{self-trapping} insulating state in the
strong-coupling adiabatic regime, no matter how many polarons it
has. The only instability which might occur in this regime is the
formation of self-trapped bipolarons, if the on-site attractive
interaction, $2\lambda zt$, is larger than the repulsive Hubbard $U$
\cite{pwa}. Self-trapped on-site bipolarons
\index{bipolaron!on-site} form a charge ordered state due to a weak
repulsion between them \cite{aleran,aub} (see also \cite{aubry}).

The transition into the self-trapped state due to a broken
translational symmetry is expected at $0.5<\lambda <1.3$ (depending
on the lattice dimensionality) for any electron-phonon interaction
conserving the on-site electron occupation numbers. For example,
Hiramoto and Toyozawa \cite{toy} calculated the strength of the
deformation potential, which transforms electrons into small
polarons and bipolarons. They found that the transition of two
electrons into a self-trapped small bipolaron occurs at the
electron-acoustic phonon coupling $\lambda \simeq 0.5$, that is half
of the critical value of $\lambda $ at which the transition of the
electron into
the small acoustic polaron takes place in the extreme adiabatic limit, $%
\omega<<zt$ (here $z$ is the coordination lattice number). The
effect of the adiabatic ratio $\omega/zt$ on the critical value of
$\lambda $ was found to be negligible. The radius of the acoustic
polaron and bipolaron is about the lattice constant, so that the
critical value of $\lambda $ does not very much depend on the number
of electrons in this case either. As discussed below the
non-adiabatic corrections (phonons) allow polarons and bipolarons to
propagate as the Bloch states in  narrow bands.

\section{Polaron dynamics}

\subsection{Polaron band}
A self-consistent approach to the multy-polaron problem is possible
with the $"1/\lambda $" expansion technique \cite{alemulty},
\index{expansion technique} which treats the kinetic energy as a
perturbation. The technique is based on the fact, known for a long
time, that there is an analytical exact solution of a $single$
polaron problem in the strong-coupling limit $\lambda \rightarrow
\infty $ \cite{lang1962a}. Following Lang and Firsov we apply the
canonical transformation $e^{S}$ to diagonalise the Hamiltonian.
\index{Lang-Firsov transformation} The diagonalisation is exact, if
$T({\bf m})=0$ (or $\lambda =\infty $):
\begin{equation}
\tilde{H}=e^{S}He^{-S},
\end{equation}
where
\begin{equation}
S=-\sum_{{\bf q},\nu ,i}\hat{n}_{i}\left[ u_{i}({\bf q,}\nu )d_{{\bf
q}\nu }-H.c.\right]
\end{equation}
is such that $S^{\dagger }=-S.$ The electron and phonon operators
are transformed as
\begin{equation}
\tilde{c}_{i}=c_{i}\exp \left[ \sum_{{\bf q}}u_{i}({\bf q,}\nu )d_{{\bf q}%
\nu }-H.c.\right] ,
\end{equation}
and
\begin{equation}
\tilde{d}_{{\bf q\nu }}=d_{{\bf q\nu }}-\sum_{i}\hat{n}_{i}u_{i}^{\ast }(%
{\bf q,}\nu ),
\end{equation}
respectively. It follows from Eq.(27) that the Lang-Firsov canonical
transformation shifts the ions to new equilibrium positions. In a
more general sense it changes the boson vacuum. As a result, the
transformed Hamiltonian takes the following form
\begin{equation}
\tilde{H}=\sum_{i,j}[\hat{\sigma}_{ij}-\mu \delta
_{ij}]c_{i}^{\dagger
}c_{j}-E_{p}\sum_{i}\hat{n}_{i}+\sum_{{\bf q,\nu }}\omega _{{\bf q\nu }}(d_{%
{\bf q\nu }}^{\dagger }d_{{\bf q\nu
}}+1/2)+{\frac{1}{{2}}}\sum_{i\neq j}v_{ij}\hat{n}_{i}\hat{n}_{j},
\end{equation}
where
\begin{equation}
\hat{\sigma}_{ij}=T({\bf m-n})\delta _{ss^{\prime }}\exp \left( \sum_{{\bf %
q,\nu }}[u_{j}({\bf q,}\nu )-u_{i}({\bf q,}\nu )]d_{{\bf q}\nu
}-H.c.\right)
\end{equation}
is the renormalised hopping integral depending on the phonon
operators, and
\begin{eqnarray}
v_{ij} &\equiv &v({\bf m-n)=} \\
&&V_{c}({\bf m-n)}-{\frac{1}{{N}}}\sum_{{\bf q,\nu }}|\gamma ({\bf
q,}\nu )|^{2}\omega _{{\bf q\nu }}\cos [{\bf q\cdot (m-n})]
\nonumber
\end{eqnarray}
is the interaction of polarons comprising their Coulomb repulsion
and the interaction via a local lattice deformation.
\index{polaron!-polaron interaction} In the extreme
infinite-coupling limit, $\lambda \rightarrow \infty ,$ we can
neglect the hopping term of the transformed Hamiltonian. The rest
has analytically determined eigenstates and eigenvalues. The
eigenstates $|\tilde{N}\rangle =|n_{i},n_{{\bf q\nu }}\rangle $ are
sorted by the polaron $n_{{\bf m}s}$ and phonon $n_{{\bf q\nu }}$
occupation numbers. The energy levels are
\begin{equation}
E=-(\mu +E_{p})\sum_{i}n_{i}+{\frac{1}{{2}}}\sum_{i\neq
j}v_{ij}n_{i}n_{j}+\sum_{{\bf q}}\omega _{{\bf q\nu }}(n_{{\bf q\nu
}}+1/2),
\end{equation}
where $n_{i}=0,1$ and $n_{{\bf q\nu }}=0,1,2,3,....\infty $.

The Hamiltonian, Eq.(27), in zero order with respect to the hopping
describes localised polarons and independent phonons, which are
vibrations of ions relative to new equilibrium positions depending
on the polaron occupation numbers. The phonon frequencies remain
unchanged in this limit. The middle of the electron band falls by
the polaron level shift $E_{p}$ due to a potential well created by
lattice deformation, Fig.2, \index{polaron!level shift}
\begin{equation}
E_{p}={\frac{1}{{2N}}}\sum_{{\bf q,\nu }}|\gamma ({\bf q,}\nu )|^{2}\omega _{%
{\bf q\nu }}.
\end{equation}

Now let us discuss the $1/\lambda $ expansion. First we restrict the
discussion to a single-polaron problem with no polaron-polaron
interaction and $\mu =0.$ The finite hopping term leads to the
polaron tunnelling because of degeneracy of the zero order
Hamiltonian with respect to the site position of the polaron. To see
how the tunnelling occurs we apply the perturbation theory using
$1/\lambda $ as a small parameter, where
\begin{equation}
\lambda \equiv \frac{E_{p}}{D},
\end{equation}
and $D=zt$. The proper Bloch set of $N$-degenerate zero order
eigenstates with the lowest energy ($-E_{p}$) of the unperturbed
Hamiltonian is
\begin{equation}
|{\bf k},0\rangle ={\frac{1}{\sqrt{N}}}\sum_{{\bf m}}c_{{\bf
m}s}^{\dagger }\exp (i{\bf k\cdot m})|0\rangle ,
\end{equation}
where $|0\rangle $ is the vacuum. By applying the textbook
perturbation theory one readily calculates the perturbed energy
levels. Up to the second order in the hopping integral they are
given by
\begin{eqnarray}
E({\bf k})=-E_{p}+\epsilon _{{\bf k}} && \\
- &&\sum_{{\bf k^{\prime }},n_{{\bf q\nu }}}{\frac{|\langle {\bf k}%
,0|\sum_{i,j}\hat{\sigma}_{ij}c_{i}^{\dagger }c_{j}|{\bf k^{\prime }},n_{%
{\bf q\nu }}\rangle |^{2}}{{\sum_{{\bf q,\nu }}\omega _{{\bf q\nu }}n_{{\bf %
q\nu }}}},}  \nonumber
\end{eqnarray}
where $|{\bf k^{\prime }},n_{{\bf q\nu }}\rangle $ are the exited
states of the unperturbed Hamiltonian with one electron and at least
one real phonon. The second term in Eq.(34), which is linear with
respect to the bare hopping $T({\bf m})$, describes the polaron-band
dispersion,
\begin{equation}
\epsilon _{{\bf k}}=\sum_{{\bf m}}T({\bf m})\,e^{-g^{2}({\bf m})}\exp (-i%
{\bf k\cdot m}),
\end{equation}
where
\begin{equation}
g^{2}({\bf m})={\frac{1}{{2N}}}\sum_{{\bf q,\nu }}|\gamma ({\bf
q,}\nu )|^{2}[1-\cos ({\bf q\cdot m})]
\end{equation}
is the {\it band-narrowing factor} at zero temperature. The third
term in
Eq.(34), quadratic in $T({\bf m})\,$, yields a negative almost ${\bf k}$-{\em %
independent} correction to the polaron level shift of the order of $%
1/\lambda ^{2}$. The origin of this correction, which could be much
larger than the fist-order contribution (Eq.(35) contains a small
exponent), is understood in Fig.2. The polaron localised in the
potential well of the depth $E_{p}$ on the site ${\bf m,}$ hops onto
a neighbouring site ${\bf n}$ with no deformation around and comes
back. As any second order correction this transition shifts the
energy down by an amount of about $-t^2/E_{p}$. It has little to do
with the polaron effective mass and the polaron tunneling mobility
because the lattice deformation around ${\bf m}$ does not follow the
electron. The electron hops back and forth many times (about
$e^{g^{2}}$) waiting for a sufficient lattice deformation to appear
around the site ${\bf n}$. Only after the deformation around ${\bf
n}$ is created does the polaron tunnel onto the next site together
with the deformation.
\begin{figure}[tbp]
\begin{center}
\includegraphics[angle=-90,width=0.75\textwidth]{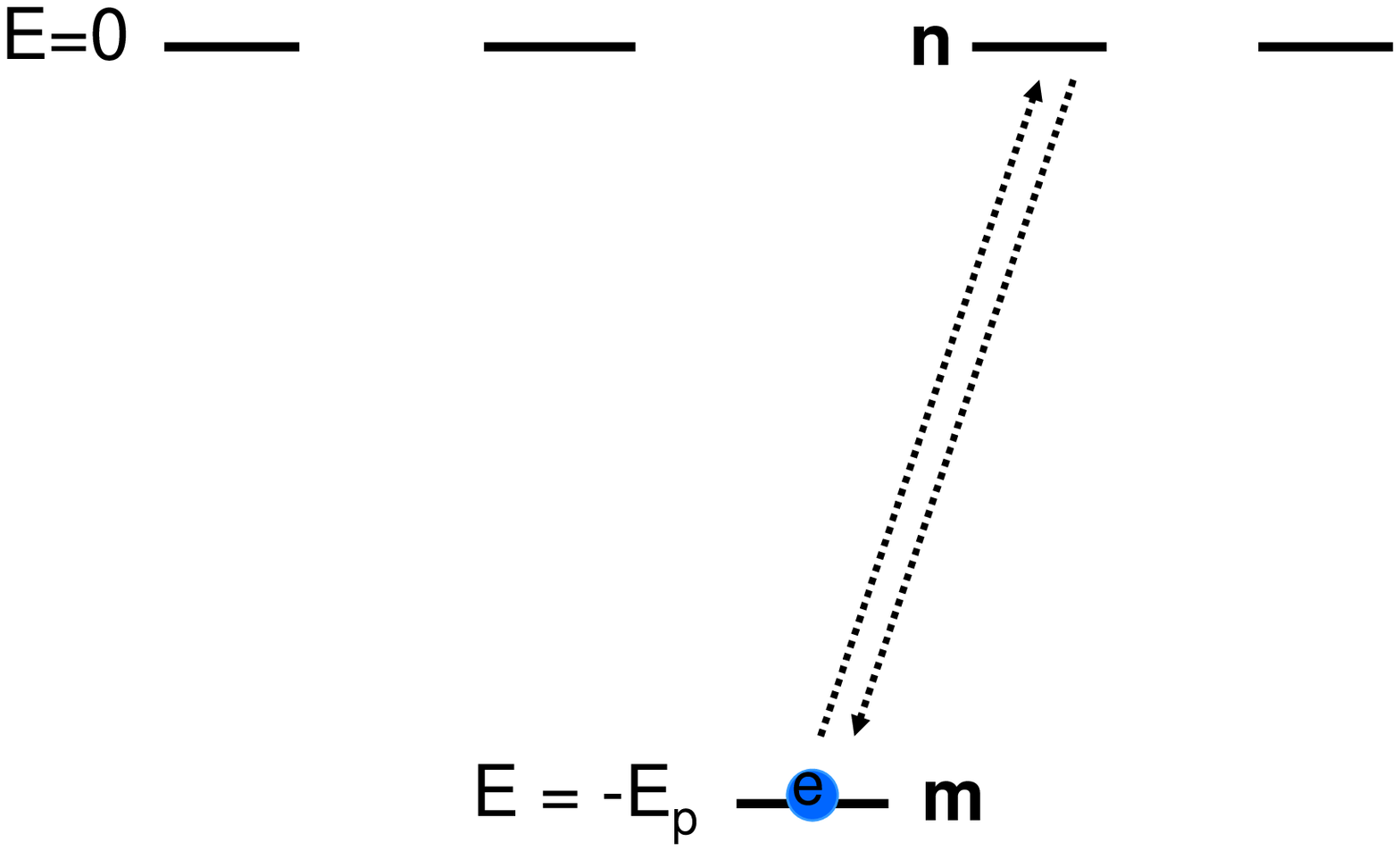}
\end{center}
\caption{"Back and forth" virtual transitions of the polaron without
any transfer of the lattice deformation from one site to another.
These transitions shift the middle of the band further down without
any real charge delocalization}
\end{figure}

Oddly enough, analysing  the Holstein two-site model  some authors
  took the second-order correction in $t$ in Eq.(34)
 as a measure of the polaron
  motion and
  arrived at
 an erroneous conclusion that "polarons are no longer describable in terms of quasiparticles having a well-defined
 dispersion" \cite{ran0} and
  "the Lang-Firsov approach, which is generally believed to become
exact in the limit of antiadiabaticity and an electron-phonon
coupling going to infinity, actually diverges  most from the exact
results precisely in this limit..." \cite{ran,ran2}. Later on it
became clear \cite{kudfir,fkka,aledyn} that this
 controversy  is the result of
the erroneous identification of the polaron kinetic energy by the
authors of Refs. \cite{ran0,ran,ran2}. \index{polaron!kinetic
energy}

\subsection{Polaron spectral and Green's functions \index{polaron!spectral function}}

The multi-polaron problem has an exact solution in the extreme
infinite-coupling limit, $\lambda =\infty ,$ for any type of e-ph
interaction conserving the on-site occupation numbers of electrons.
 For the finite coupling $1/\lambda $ perturbation
expansion is applied. \index{multi-polaron problem!exact solution}
The expansion parameter is actually \cite{fir,eag,gog,alemulty}
\[
\frac{1}{2z\lambda ^{2}}\ll 1,
\]
so that the analytical perturbation theory has a wider region of
applicability than one can expect using a semiclassical estimate
$E_{p}>D$. However, the expansion convergency is different for
different e-ph interactions. Exact numerical diagonalisations of
vibrating clusters, variational calculations (see Refs.
\cite{kab2,feh,mar,feh,rom,trugman,wag} and \cite{cataud,fehske}),
\index{variational approach} dynamical mean-field approach in
infinite dimensions \cite{zey}, \index{dynamical mean-field theory}
and Quantum-Monte-Carlo
 simulations (see Refs. \cite{Raedt,Korn2,Hoh,pro,Mac,spen,hag,jim} and \cite{kornil,nagaosa})
 \index{Quantum-Monte-Carlo algorithm}
 simulations revealed
that the ground state energy ($\approx -$ $E_{p})$ is not very
sensitive to the parameters. On the contrary, the effective mass,
the bandwidth, and the polaron density of states strongly depend on
the adiabatic ratio $\omega /t$ and on the radius of the
interaction. The first-order in $1/\lambda $ perturbation theory is
practically exact in the non-adiabatic regime $\omega > t$
\index{non-adiabatic regime} $for$ $any$ $value$ of the coupling
constant and any type of e-ph interaction. However, it
$overestimates$ the polaron mass by a few orders of magnitude in the
adiabatic case, $\omega \ll t,$ if the interaction is short-ranged
\cite{kab2}.

A much lower effective mass of the adiabatic Holstein polaron
\index{polaron!Holstein} compared with that estimated using the
first order perturbation theory is the result of poor convergency of
the perturbation expansion owing to the double-well potential
\cite{hol} in the adiabatic limit. The tunnelling probability is
extremely sensitive to the shape of this potential and also to the
phonon frequency dispersion. The latter leads to a much lighter
Holstein polaron compared with the nondispersive approximation
\cite{zoli}. \index{Holstein!model!dispersive} Importantly, the
analytical perturbation theory becomes practically exact in a wider
range of the adiabatic parameter and of the coupling constant for
the long-range Fr\"{o}hlich interaction \cite{Korn2}.
\index{Fr\"ohlich!interaction}

Keeping this in mind, let us calculate the one-particle GF in the
first order in $1/\lambda $. \index{polaron!Green's function}
Applying the canonical transformation we write the transformed
Hamiltonian as
\begin{equation}
\tilde{H}=H_{p}+H_{ph}+H_{int},
\end{equation}
where
\begin{equation}
H_{p}=\sum_{{\bf k}}\xi ({\bf k})c_{{\bf k}}^{\dagger }c_{{\bf k}}
\end{equation}
is the ``free'' $polaron$ contribution,
\begin{equation}
H_{ph}=\sum_{{\bf q}}\omega _{{\bf q}}(d_{{\bf q}}^{\dagger }d_{{\bf
q}}+1/2)
\end{equation}
is the free phonon part (spin and phonon branch quantum numbers are
dropped here), and $\xi_{\bf k}=Z^{\prime }E_{\bf k}-\mu $ is the
renormalised polaron-band dispersion. The chemical potential $\mu $
includes the polaron level shift $-E_{p}$, and it could also include
all higher orders in $1/\lambda $ corrections to the polaron
spectrum, which are independent of ${\bf k}$. The band-narrowing
factor $Z^{\prime}$ is defined as
\begin{equation}
Z^{\prime }={\frac{\sum_{{\bf m}}T({\bf m})e^{-g^{2}({\bf m})}\exp (-i{\bf %
k\cdot m})}{{\sum_{{\bf m}}T({\bf m})\exp (-i{\bf k\cdot m})}}},
\end{equation}
which is $Z^{\prime }=\exp (-\gamma E_{p}/\omega ))$ with
$\gamma\leq 1$ depending on the range of the e-ph interaction and
phonon frequency dispersions. The interaction term $H_{int}$
comprises the polaron-polaron interaction, Eq.(29), and the residual
polaron-phonon interaction
\begin{equation}
H_{p-ph}\equiv \sum_{i\neq ,j}[\hat{\sigma}_{ij}-\left\langle \hat{\sigma}%
_{ij}\right\rangle _{ph}]c_{i}^{\dagger }c_{j},
\end{equation}
where $\left\langle \hat{\sigma}%
_{ij}\right\rangle _{ph}$ means averaging with respect to the bare
phonon distribution. We can neglect $H_{p-ph}$ in the first-order of
$1/\lambda \ll 1$. To understand spectral properties of a single
polaron we also neglect the polaron-polaron interaction. Then the
energy levels are
\begin{equation}
E_{\tilde{m}}=\sum_{_{{\bf k}}}\xi _{_{{\bf k}}}n_{{\bf k}}+\sum_{{\bf q}%
}\omega _{{\bf q}}[n_{{\bf q}}+1/2],
\end{equation}
and the transformed eigenstates $\left| \tilde{m}\right\rangle $ are
sorted by the polaron Bloch-state occupation numbers, $n_{{\bf
k}}=0,1$, and the phonon occupation numbers, $n_{{\bf
q}}=0,1,2,...,\infty .$

The spectral function of any system described by  quantum numbers
$m,n$ and eigenvalues $E_n, E_m$ is defined as (see, for example
\cite{alebook})
\begin{equation}
A({\bf k},\omega )\equiv \pi (1+e^{-\omega /T})e^{\Omega
/T}\sum_{n,m}e^{-E_{n}/T}\left| \left\langle n\right| c_{{\bf
k}}\left| m\right\rangle \right| ^{2}\delta (\omega _{nm}+\omega ).
\end{equation}
It is real and  positive, $A({\bf k},\omega )>0$, and obeys the
important sum rule
\begin{equation}
\frac{1}{\pi }\int_{-\infty }^{\infty }d\omega A({\bf k},\omega )=1.
\end{equation}
Here $\Omega$ is the thermodynamic potential. The matrix elements of
the electron operators can be written as
\begin{equation}
\left\langle n\right| c_{{\bf k}}\left| m\right\rangle =\frac{1}{\sqrt{N}}%
\sum_{{\bf m}}e^{-i{\bf k\cdot m}}\left\langle \tilde{n}\right| c_{i}\hat{X}%
_{i}\left| \tilde{m}\right\rangle
\end{equation}
by the use of the Wannier representation and the Lang-Firsov
transformation. Here
\[
\hat{X}_{i}=\exp \left[ \sum_{{\bf q}}u_{i}({\bf q})d_{{\bf
q}}-H.c.\right] .
\]
Now, applying the Fourier transform of the $\delta $-function in
Eq.(43),
\[
\delta (\omega _{nm}+\omega )=\frac{1}{2\pi }\int_{-\infty }^{\infty
}dte^{i(\omega _{nm}+\omega )t},
\]
the spectral function \index{polaron!spectral function} is expressed
as
\begin{eqnarray}
A({\bf k},\omega ) &=&\frac{1}{2}\int_{-\infty }^{\infty }dte^{i\omega t}%
\frac{1}{N}\sum_{{\bf m,n}}e^{i{\bf k\cdot (n-m)}}\times \\
&&\left\{ \left\langle \left\langle c_{i}(t)\hat{X}_{i}(t)c_{j}^{\dagger }%
\hat{X}_{j}^{\dagger }\right\rangle \right\rangle +\left\langle
\left\langle c_{j}^{\dagger }\hat{X}_{j}^{\dagger
}c_{i}(t)\hat{X}_{i}(t)\right\rangle \right\rangle \right\} .
\nonumber
\end{eqnarray}
Here the quantum and statistical averages are performed for
independent polarons and phonons, therefore
\begin{equation}
\left\langle \left\langle c_{i}(t)\hat{X}_{i}(t)\hat{X}_{j}^{\dagger
}c_{i}^{\dagger }\right\rangle \right\rangle =\left\langle
\left\langle c_{i}(t)c_{j}^{\dagger }\right\rangle \right\rangle
\left\langle \left\langle \hat{X}_{i}(t)\hat{X}_{j}^{\dagger
}\right\rangle \right\rangle .
\end{equation}
The Heisenberg free-polaron operator evolves with time as
\begin{equation}
c_{{\bf k}}(t)=c_{{\bf k}}e^{-i\xi _{{\bf k}}t},
\end{equation}
and
\begin{eqnarray}
\left\langle \left\langle c_{i}(t)c_{i}^{\dagger }\right\rangle
\right\rangle &=&\frac{1}{N}\sum_{{\bf k}^{\prime }{\bf ,k}^{\prime
\prime
}}e^{i({\bf k}^{\prime }{\bf \cdot m-k}^{\prime \prime }{\bf \cdot n)}%
}\left\langle \left\langle c_{{\bf k}^{\prime }}(t)c_{{\bf
k}^{\prime \prime
}}^{\dagger }\right\rangle \right\rangle = \\
&&\frac{1}{N}\sum_{{\bf k}^{\prime }}[1-\bar{n}({\bf k}^{\prime })]e^{i{\bf k%
}^{\prime }{\bf \cdot (m-n)-}i\xi _{{\bf k}^{\prime }}t},  \nonumber \\
\left\langle \left\langle c_{i}^{\dagger }c_{i}(t)\right\rangle
\right\rangle &=&\frac{1}{N}\sum_{{\bf k}^{\prime }}\bar{n}({\bf
k}^{\prime })e^{i{\bf k}^{\prime }{\bf \cdot (m-n)-}i\xi _{{\bf
k}^{\prime }}t}
\end{eqnarray}
where $\bar{n}({\bf k)}=[1+\exp \xi _{{\bf k}}/T]^{-1}$ is the
Fermi-Dirac distribution function of polarons. The Heisenberg
free-phonon operator evolves in a similar way,
\[
d_{{\bf q}}(t)=d_{{\bf q}}e^{-i\omega _{{\bf q}}t},
\]
and
\begin{equation}
\left\langle \left\langle \hat{X}_{i}(t)\hat{X}_{j}^{\dagger
}\right\rangle
\right\rangle =\prod_{_{{\bf q}}}\left\langle \left\langle \exp [u_{i}({\bf %
q,}t)d_{{\bf q}}-H.c.]\exp [-u_{j}({\bf q})d_{{\bf
q}}-H.c.]\right\rangle \right\rangle ,
\end{equation}
where $u_{i,j}({\bf q,}t)=u_{i,j}({\bf q})e^{-i\omega _{{\bf q}}t}.$
This average is calculated using the operator identity
\begin{equation}
e^{\hat{A}+\hat{B}}=e^{\hat{A}}e^{\hat{B}}e^{-[\hat{A},\hat{B}]/2},
\end{equation}
which is applied for any two operators $\hat{A}$ and $\hat{B}$,
whose
commutator $[\hat{A},\hat{B}]$ is a number. Because $[d_{{\bf q}},d_{{\bf q}%
}^{\dagger }]=1,$ we can apply this identity in Eq.(51) to obtain
\begin{eqnarray*}
e^{[u_{i}({\bf q,}t)d_{{\bf q}}-H.c.]}e^{[-u_{j}({\bf q})d_{{\bf
q}}-H.c.]}
&=&e^{(\alpha ^{\ast }d_{{\bf q}}^{\dagger }-\alpha d_{{\bf q}})}\times \\
&&e^{[u_{i}({\bf q,}t)u_{j}^{\ast }({\bf q})-u_{i}^{\ast }({\bf q,}t)u_{j}(%
{\bf q})]/2},
\end{eqnarray*}
where $\alpha \equiv u_{j}({\bf q,}t)-u_{i}({\bf q}).$ Applying once
again the same identity yields
\begin{eqnarray}
e^{[u_{i}({\bf q,}t)d_{{\bf q}}-H.c.]}e^{[-u_{j}({\bf q})d_{{\bf
q}}-H.c.]}
&=&e^{\alpha ^{\ast }d_{{\bf q}}^{\dagger }}e^{-\alpha d_{{\bf q}%
}}e^{-|\alpha |^{2}/2}\times \\
&&e^{[u_{i}({\bf q,}t)u_{j}^{\ast }({\bf q})-u_{i}^{\ast }({\bf q,}t)u_{j}(%
{\bf q})]/2}.  \nonumber
\end{eqnarray}
Now the quantum and statistical averages are calculated by the use
of
\begin{equation}
\left\langle \left\langle e^{\alpha ^{\ast }d_{{\bf q}}^{\dagger
}}e^{-\alpha d_{{\bf q}}}\right\rangle \right\rangle =e^{-|\alpha
|^{2}n_{\omega }},
\end{equation}
where $n_{\omega }=[\exp (\omega _{{\bf q}}/T)-1]^{-1}$ is the
Bose-Einstein distribution function of phonons. Collecting all
multiplies in Eq.(51) we arrive at
\begin{equation}
\left\langle \left\langle \hat{X}_{i}(t)\hat{X}_{j}^{\dagger
}\right\rangle
\right\rangle =\exp \left\{ -\frac{1}{2N}\sum_{{\bf q}}|\gamma ({\bf q}%
)|^{2}f_{{\bf q}}({\bf m-n,}t)\right\} ,
\end{equation}
where
\begin{equation}
f_{{\bf q}}({\bf m,}t)=[1-\cos ({\bf q\cdot m)}\cos (\omega _{{\bf q}%
}t)]\coth \frac{\omega _{{\bf q}}}{2T}+i\cos ({\bf q\cdot m)}\sin (\omega _{%
{\bf q}}t).
\end{equation}
Here we used the symmetry of $\gamma (-{\bf q})=\gamma ({\bf q})$,
because of which  terms containing $\sin ({\bf q\cdot m)}$
disappeared. The
average $\left\langle \left\langle \hat{X}_{j}^{\dagger }\hat{X}%
_{i}(t)\right\rangle \right\rangle ,$ which is a multiplier in the
second term in the brackets of Eq.(46), is obtained as
\begin{equation}
\left\langle \left\langle \hat{X}_{j}^{\dagger
}\hat{X}_{i}(t)\right\rangle \right\rangle =\left\langle
\left\langle \hat{X}_{i}(t)\hat{X}_{j}^{\dagger }\right\rangle
\right\rangle ^{\ast }.
\end{equation}
To proceed with the analytical results we consider low temperatures,
$T\ll \omega _{{\bf q}},$ when $\coth (\omega _{{\bf q}}/2T)\approx
1.$ Then expanding the exponent in Eq.(55) yields
\begin{equation}
\left\langle \left\langle \hat{X}_{i}(t)\hat{X}_{j}^{\dagger
}\right\rangle
\right\rangle =Z\sum_{l=0}^{\infty }\frac{\left\{ \sum_{{\bf q}}|\gamma (%
{\bf q})|^{2}e^{i[{\bf q\cdot (m-n)-}\omega _{{\bf q}}t]}\right\} ^{l}}{%
(2N)^{l}l!},
\end{equation}
where
\begin{equation}
Z=\exp \left[ -{\frac{1}{2N}}\sum_{{\bf q}}|\gamma ({\bf
q})|^{2}\right] .
\end{equation}
Then performing summation over ${\bf m}, {\bf n}$, ${\bf k}^{\prime
}$ and integration over time in Eq.(46) we arrive at \cite{alechan}
\begin{equation}
A({\bf k},\omega )=\sum_{l=0}^{\infty }\left[ A_{l}^{(-)}({\bf
k},\omega )+A_{l}^{(+)}({\bf k},\omega )\right] ,
\end{equation}
where
\begin{eqnarray}
A_{l}^{(-)}({\bf k},\omega ) &=&\pi Z\sum_{{\bf q}_{1},...{\bf q}_{l}}{\frac{%
\prod_{r=1}^{l}|\gamma ({\bf q}_{r})|^{2}}{{(2N)^{l}l!}}\times } \\
&&\left[ 1-\bar{n}\left( {{\bf k-}\sum_{r=1}^{l}{\bf q}_{r}}\right) \right] {%
\delta }\left( {\omega -}\sum_{r=1}^{l}\omega _{{\bf q}_{r}}{-\xi }_{{{\bf k-%
}\sum_{r=1}^{l}{\bf q}_{r}}}\right) ,  \nonumber
\end{eqnarray}
and
\begin{eqnarray}
A_{l}^{(+)}({\bf k},\omega ) &=&\pi Z\sum_{{\bf q}_{1},...{\bf q}_{l}}{\frac{%
\prod_{r=1}^{l}|\gamma ({\bf q}_{r})|^{2}}{{(2N)^{l}l!}}\times } \\
&&\bar{n}\left( {{\bf k+}\sum_{r=1}^{l}{\bf q}_{r}}\right) {\delta }\left( {%
\omega +}\sum_{r=1}^{l}\omega _{{\bf q}_{r}}{-\xi }_{{{\bf k+}\sum_{r=1}^{l}%
{\bf q}_{r}}}\right) .  \nonumber
\end{eqnarray}
Clearly Eq.(60) is in the form of a perturbative multi-phonon
expansion. Each contribution $A_{l}^{(\pm )}({\bf k},\omega )$ to
the spectral function describes the transition from the initial
state ${\bf k}$
of the polaron band to the final state ${\bf k}{\bf \pm }$ ${\sum_{r=1}^{l}%
{\bf q}_{r}}$ with the emission (or absorption) of $l$ phonons.

The $%
1/\lambda $ expansion result \index{expansion technique}, Eq.(60),
is applied to {\it low-energy} polaron excitations in the
strong-coupling limit. In the case of the long-range Fr\"{o}hlich
interaction with high-frequency phonons it is also applied in the
intermediate regime \cite{Korn2,hag}. Different from the
conventional ME spectral function there is no imaginary part of the
self-energy since the exponentially small at low temperatures
polaronic damping \cite{alebook} is neglected. Instead the e-ph
coupling leads to the coherent dressing of electrons by phonons. The
dressing can be seen as the phonon "side-bands" with $l\geq 1.$
\index{phonon!side-bands}

While the major sum rule, Eq.(44) is satisfied, \index{sum rule}
\begin{eqnarray}
\frac{1}{\pi }\int_{-\infty }^{\infty }d\omega A({\bf k},\omega )
&=&Z\sum_{l=0}^{\infty }\sum_{{\bf q}_{1},...{\bf q}_{l}}{\frac{%
\prod_{r=1}^{l}|\gamma ({\bf q}_{r})|^{2}}{{(2N)^{l}l!}}=} \\
Z\sum_{l=0}^{\infty }\frac{1}{{l!}}\left\{ \frac{1}{2N}{\sum_{{\bf q}%
}|\gamma ({\bf q})|^{2}}\right\} ^{l} &=&Z\exp \left[ {\frac{1}{2N}}\sum_{%
{\bf q}}|\gamma ({\bf q})|^{2}\right] =1,  \nonumber
\end{eqnarray}
the higher-momentum integrals, $\int_{-\infty }^{\infty }d\omega
\omega ^{p}A({\bf k},\omega )$ with $p>0,$ calculated using Eq.(60),
differ from the exact values (see Part III)  by an amount
proportional to $1/\lambda .$  The difference is due to a partial
"undressing" of high-energy excitations in the side-bands, which is
beyond the first order $1/\lambda $ expansion.

The spectral function of the polaronic carriers comprises two
different parts. The first ($l=0$) ${\bf k}$-dependent {\it
coherent} term arises from the polaron band tunnelling,
\begin{equation}
A_{coh}({\bf k},\omega )=\left[ A_{0}^{(-)}({\bf k},\omega )+A_{0}^{(+)}(%
{\bf k},\omega )\right] =\pi {Z\delta (\omega -\xi }_{{\bf k}}{)}.
\end{equation}
The spectral weight of the coherent part is suppressed as $Z\ll 1.$
However in the case of the Fr\"{o}hlich interaction the effective
mass is less
enhanced, $\xi _{{\bf k}}=Z^{\prime }E_{{\bf k}}-\mu $, because $%
Z<<Z^{\prime }<1$.

The second {\it incoherent} part $%
A_{incoh}({\bf k},\omega )$ comprises all the terms with $l\geq 1.$
It describes the excitations accompanied by emission and absorption
of phonons. We notice that its spectral density spreads over a wide
energy range of about twice the polaron level shift $E_{p}$, which
might be larger than the unrenormalised bandwidth $2D$ in the rigid
lattice without phonons. On the contrary, the coherent part shows a
dispersion only in the energy window of the order of the polaron
bandwidth, $w=Z^{\prime }D$. It is interesting that there is some
${\bf k}$ dependence of the $incoherent$ background as well
\cite{alechan}, if the matrix element of the e-ph interaction and/or
phonon frequencies depend on ${\bf q}$. Only in the Holstein model
with the short-range
dispersionless e-ph interaction $\gamma ({\bf q)=}\gamma _{0}$ and $\omega _{%
{\bf q}}=\omega _{0}$ the incoherent part is momentum independent
(see also Ref. \cite{mahan}),
\begin{eqnarray}
A_{incoh}({\bf k},\omega ) &=&\pi \frac{{Z}}{N}\sum_{l=1}^{\infty }{\frac{%
\gamma _{0}^{2l}}{{2}^{l}{l!}}\times } \\
&&\sum_{{\bf k}^{\prime }}\left\{ \left[ 1-\bar{n}\left( {\bf
k}^{\prime
}\right) \right] {\delta }\left( {\omega -}l\omega _{0}{-\xi }_{{\bf k}%
^{\prime }}\right) +\bar{n}\left( {\bf k}^{\prime }\right) {\delta }\left( {%
\omega +}l\omega _{0}{-\xi }_{{\bf k}^{\prime }}\right) \right\} .
\nonumber
\end{eqnarray}

As soon as we know the spectral function, different Green's
functions (GF) are readily obtained using their analytical
properties. index{polaron!Green's function} For example, the
temperature GF is given by the integral \cite{alebook}
\begin{equation}
{\cal G}({\bf k},\omega _{k})=\frac{1}{\pi }\int_{-\infty }^{\infty
}d\omega
^{\prime }\frac{A({\bf k},\omega ^{\prime })}{i\omega _{k}-\omega ^{\prime }}%
.
\end{equation}
where $\omega _{k}=\pi T(2k+1),k=0,\pm 1,\pm 2,...$. Calculating the
integral with the spectral density Eq.(65) we find in the Holstein
model \cite{aleranG}
\begin{equation}
{\cal G}({\bf k},\omega _{n})=\frac{Z}{i{\omega }_{n}{-\xi }_{{\bf k}}}+%
\frac{{Z}}{N}\sum_{l=1}^{\infty }{\frac{\gamma _{0}^{2l}}{{2}^{l}{l!}}}\sum_{%
{\bf k}^{\prime }}\left\{ \frac{1-\bar{n}\left( {\bf k}^{\prime }\right) }{i{%
\omega }_{n}{-}l\omega _{0}{-\xi }_{{\bf k}^{\prime
}}}+\frac{\bar{n}\left(
{\bf k}^{\prime }\right) }{i{\omega }_{n}{+}l\omega _{0}{-\xi }_{{\bf k}%
^{\prime }}}\right\} .
\end{equation}
Here the first term describes the coherent tunnelling in the narrow
polaron band while the second ${\bf k}$-independent sum describes
the phonon-side bands.

\section{Polaron-polaron interaction \index{polaron!-polaron interaction}and small bipolaron \index{bipolaron!small}}

Polarons interact with each other, Eq.(29). The range of the
deformation surrounding the Fr\"{o}hlich polarons
\index{polaron!Fr\"{o}hlich} is quite large, and their deformation
fields are overlapped at finite density. Taking into account both
the long-range attraction of polarons owing to the lattice
deformations $and$ their direct Coulomb repulsion, the residual {\it
long-range} interaction turns out rather weak and repulsive in ionic
crystals \cite{alemot}. The Fourier component of the polaron-polaron
interaction, $v({\bf q})$, comprising the direct Coulomb repulsion
and the attraction mediated by phonons is
\begin{equation}
v({\bf q})={\frac{4\pi e^{2}}{{\epsilon _{\infty }q^{2}}}}-|\gamma ({\bf q}%
)|^{2}\omega _{{\bf q}}.
\end{equation}
In the long-wave limit ($q \ll \pi /a$) the Fr\"{o}hlich interaction
\cite{fro2} dominates in the attractive part, which is described by
Eq.(9). Fourier transforming Eq.(68) yields the repulsive
interaction in real space,
\begin{equation}
v({\bf m-n)}={\frac{e^{2}}{{\epsilon _{0}|{\bf m-n}|}}>0}.
\end{equation}
We see that optical phonons nearly nullify the bare Coulomb
repulsion in ionic solids, where $\epsilon _{0}\gg 1$, but cannot
overscreen it at large distances.

Considering the polaron-phonon interaction \index{polaron!-phonon
interaction} in the multi-polaron system we have to take into
account dynamic properties of the polaron response function
\cite{aledyn}. One may erroneously believe that the long-range
Fr\"{o}hlich interaction becomes a short-range (Holstein) one due to
the screening of ions by heavy polaronic carriers. In fact, small
polarons cannot screen high-frequency optical vibrations because
their renormalised plasma frequency \index{polaron!plasma frequency}
is comparable with or even less than the phonon frequency. In the
absence of bipolarons (see below) we can apply the ordinary bubble
approximation  to calculate the dielectric response function of
polarons at the frequency $\Omega $ as
\begin{equation}
\epsilon ({\bf q},\Omega )=1-2v({\bf q})\sum_{{\bf k}}{\frac{\bar{n}({\bf %
k+q)}-\bar{n}({\bf k)}}{{\Omega -\epsilon _{{\bf k}}+\epsilon _{{\bf k+q}}}}}%
.
\end{equation}
This expression describes the response of small polarons to any
external field of the frequency $\Omega \leq \omega _{0}$, when
phonons in the polaron cloud follow the polaron motion. In the
static limit we obtain the usual Debye screening at large distances
($q\rightarrow 0$). For the temperature larger than the polaron
half-bandwidth, $T>w,$ we can approximate the polaron distribution
function as \index{polaron!distribution function}
\begin{equation}
\bar{n}({\bf k)}\approx \frac{n}{{2}a^{3}}\left( 1-{\frac{(2-n)\epsilon _{%
{\bf k}}}{{2T}}}\right) ,
\end{equation}
and obtain
\begin{equation}
\epsilon (q,0)=1+{\frac{q_{s}^{2}}{{q^{2}}}},
\end{equation}
where
\[
q_{s}=\left[ \frac{2\pi e^{2}n(2-n)}{\epsilon _{0}Ta^{3}}\right]
^{1/2},
\]
and $n$ is the number of polarons per  unit cell. For a finite but
rather low frequency, $\omega _{0} > \Omega \gg w,$ the polaron
response becomes dynamic,
\begin{equation}
\epsilon ({\bf q},\Omega )=1-{\frac{\omega _{p}^{2}({\bf
q})}{{\Omega ^{2}}}}
\end{equation}
where
\begin{equation}
\omega _{p}^{2}({\bf q})=2v({\bf q})\sum_{{\bf k}}n({\bf k)}(\epsilon _{{\bf %
k+q}}-\epsilon _{{\bf k}})
\end{equation}
is the temperature-dependent polaron plasma frequency squared. The
polaron plasma frequency is rather low due to the large static
dielectric constant, $\epsilon _{0}\gg 1,$ and the enhanced polaron
mass $m^{\ast }\gg m_e$.

Now replacing the bare electron-phonon interaction vertex $\gamma
({\bf q)}$ by a screened one, $\gamma _{sc}({\bf q},\omega _{0}),$
as shown in Fig. 3, we obtain
\begin{equation}
\gamma _{sc}({\bf q},\omega _{0})={\frac{\gamma ({\bf q})}{{\epsilon ({\bf q}%
,\omega }_{0}{)}}\approx }\gamma ({\bf q})
\end{equation}
because ${\omega }_{0}>\omega _{p}.$ Therefore, the singular behaviour of $%
\gamma ({\bf q})\sim 1/q$ is unaffected by screening. Polarons are
too slow to screen high-frequency crystal field oscillations. As a
result, the strong interaction with high-frequency optical phonons
in ionic solids remains unscreened at any density of small polarons.

Another important point is a possibility of the Wigner
crystallization \index{polaron!Wigner crystallization} of the
polaronic liquid. Because the net long-range repulsion is relatively
weak, a relevant dimensionless parameter $r_{s}=m^{\ast
}e^{2}/\epsilon _{0}(4\pi n/3)^{1/3}$ is not very large in ionic
semiconductors. The Wigner crystallization appears around
$r_{s}\simeq 100$ or larger, which corresponds to the atomic density
of polarons $n\leq 10^{-6}$ with $\epsilon _{0}=30$ and $m^{\ast
}=5m_{e}$. This estimate tells us that polaronic carriers are
usually in the liquid state at relevant doping levels.

\begin{figure}[tbp]
\begin{center}
\includegraphics[angle=-90,width=0.75\textwidth]{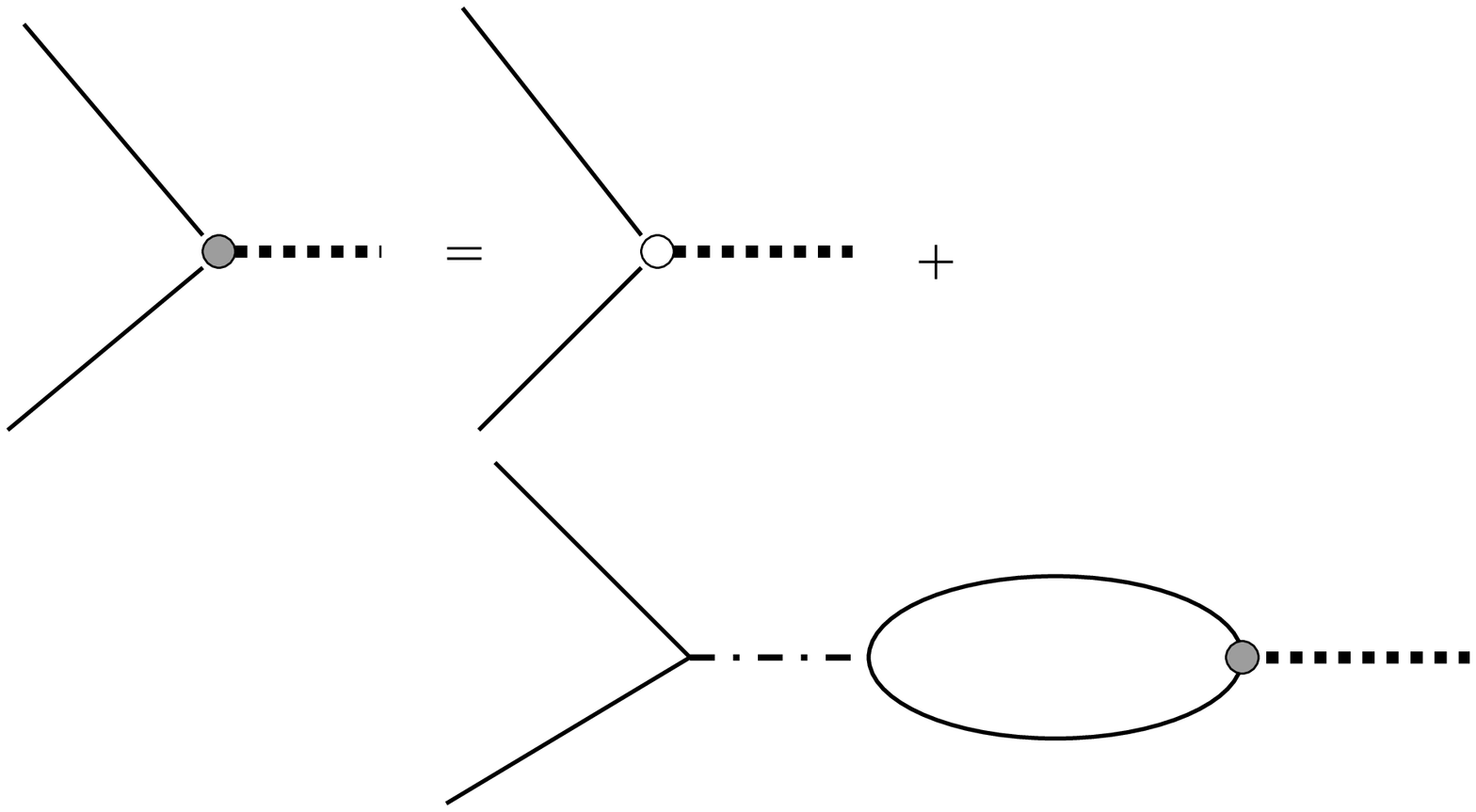}
\end{center}
\caption{E-ph vertex, $\gamma({\bf q})$ screened by the
polaron-polaron interaction, $v({\bf q})$ (dashed-dotted line).
Solid and dotted lines are polaron and phonon propagators,
respectively.}
\end{figure}

At large distances polarons repel each other. Nevertheless two $%
large$ polarons \index{polaron!large} can be bound into a $large$
bipolaron \index{bipolaron!large} by an exchange interaction even
with no additional e-ph interaction but the Fr\"{o}hlich one
\cite{vin,dev,dev0}. When a short-range deformation potential and
molecular-type (i.e. Holstein) e-ph interactions
\index{Holstein!interaction} are taken into account together with
the Fr\"{o}hlich interaction, \index{Fr\"ohlich!interaction} they
overcome the Coulomb repulsion at a short distance of about the
lattice constant. Then, owing to a narrow band, two small polarons
easily form a bound state, i.e. a $small$ bipolaron. Let us estimate
the coupling constant $\lambda $ and the adiabatic ratio $\omega
_{0}/t,$ at which the  "bipolaronic" instability occurs. The
characteristic attractive potential is $|v|=D/(\lambda -\mu _{c})$,
where $\mu _{c}$ is the dimensionless Coulomb repulsion, and
$\lambda $ includes the interaction with all phonon branches. The
radius of the potential is about $a$. In three dimensions a bound
state of two attractive particles appears, if
\begin{equation}
|v| \geq \frac{\pi ^{2}}{8m^{\ast }a^{2}}.
\end{equation}
Substituting the polaron mass, $m^{\ast }=[2a^{2}t]^{-1}\exp (\gamma
\lambda D/\omega _{0})$, we find
\begin{equation}
\frac{t}{{\omega }_{0}}\leq (\gamma z\lambda )^{-1}\ln \left[
{\frac{\pi ^{2}}{{4z(\lambda -\mu }_{c}{)}}}\right] .
\end{equation}
As a result, small bipolarons form at $\lambda \geq \mu _{c}+\pi
^{2}/4z$ almost independent of the adiabatic ratio. In the case of
the Fr\"{o}hlich interaction there is no sharp transition between
small and large polarons, and the first-order $1/\lambda $ expansion
is accurate in the whole region of the e-ph coupling, if the
adiabatic parameter is not too
small. Hence we can say that in the antiadiabatic and intermediate regime the carriers are small polarons $%
independent$ of the value of $\lambda $ if the e-ph interaction is
long-ranged.  It means that they tunnel together with the entire
phonon cloud no matter how "thin" the cloud is. \index{phonon cloud}

\section{Polaronic superconductivity \index{superconductivity!polaronic}}

The polaron-polaron interaction \index{polaron!-polaron interaction}
is the sum of two large contributions of the opposite sign, Eq.(29).
It is generally larger than the polaron bandwidth and the polaron
Fermi-energy, $\epsilon _{F}=Z^{\prime }E_{F}$. This condition is
opposite to the weak-coupling BCS regime, where the Fermi energy is
the largest. However, there is still a narrow window of parameters,
where bipolarons are extended enough, and pairs of two small
polarons are overlapped similar to the Cooper pairs. Here the BCS
approach is applied to nonadiabatic carriers with a {\it
nonretarded} attraction, so that bipolarons are the Cooper pairs
formed by two {\it small} {\it polarons} \cite{ale0}. The size of
the bipolaron is estimated as
\begin{equation}
r_{b}\approx \frac{1}{(m^{\ast }\Delta )^{1/2}},
\end{equation}
where $\Delta$ is the binding energy of the order of an attraction
potential $v.$ The BCS approach is applied if $r_{b}\gg n^{-1/3},$
which puts a severe constraint on the value of the attraction
\begin{equation}
|v|\ll \epsilon _{F}.
\end{equation}
There is no "Tolmachev" logarithm  in the case of nonadiabatic
carriers, because the attraction is nonretarded as soon as $\epsilon
_{F}\leq \omega _{0}$. Hence a superconducting state of small
polarons is possible only if $\lambda >\mu _{c}$. This consideration
leaves a rather narrow $crossover$ region from the normal polaron
Fermi liquid to a superconductor, where one can still apply the BCS
mean-field approach,
\begin{equation}
0<\lambda -\mu _{c}\ll Z^{\prime }<1.
\end{equation}
\ In the case of the Fr\"{o}hlich interaction $Z^{\prime }$ is about $%
0.1\div 0.3$ for typical values of $\lambda.$ Hence the region,
Eq.(80), is on the border-line from the polaronic normal metal to a
bipolaronic superconductor (section 8).

In the crossover region polarons behave like fermions in a narrow
band with a weak nonretarded attraction. As long as $\lambda \gg
1/\sqrt{2z}$, we can neglect their residual interaction with phonons
in the transformed Hamiltonian,
\begin{equation}
\tilde{H}\approx \sum_{i,j}\left[ (\left\langle \hat{\sigma}%
_{ij}\right\rangle _{ph}-\mu \delta _{ij})c_{i}^{\dagger }c_{j}+{\frac{1}{{2}%
}}v_{ij}c_{i}^{\dagger }c_{j}^{\dagger }c_{j}c_{i}\right]
\end{equation}
written in the Wannier representation
\index{Wannier!representation}. If the condition Eq.(80) is
satisfied, we can treat the polaron-polaron interaction
approximately by the use of the BCS theory. For simplicity let us
keep only the on-site $v_{0}$ and the nearest-neighbour $v_{1}$
interactions. At least one of them should be attractive to ensure
that the ground state is superconducting. Introducing two order
parameters
\begin{equation}
\Delta _{0}=-v_{0}\langle c_{{\bf m},\uparrow }c_{{\bf m},\downarrow
}\rangle ,
\end{equation}
\begin{equation}
\Delta _{1}=-v_{1}\langle c_{{\bf m},\uparrow }c_{{\bf
m+a},\downarrow }\rangle
\end{equation}
and transforming to the ${\bf k}$-space results in the familiar BCS
Hamiltonian,
\begin{equation}
H_{p}=\sum_{{\bf k},s}\xi _{{\bf k}}c_{{\bf k}s}^{\dagger }c_{{\bf k}%
s}+\sum_{{\bf k}}[\Delta _{{\bf k}}c_{{\bf k}\uparrow }^{\dagger }c_{{\bf -k}%
\downarrow }^{\dagger }+H.c.],
\end{equation}
where $\xi _{{\bf k}}=\epsilon _{{\bf k}}-\mu $ is the renormalised
kinetic energy and
\begin{equation}
\Delta _{{\bf k}}=\Delta _{0}-\Delta _{1}{\frac{\xi _{{\bf k}}+\mu
}{{w}}}
\end{equation}
is the order parameter.

The Bogoliubov anomalous averages \index{Bogoliubov averages} are
found as
\begin{equation}
\langle c_{{\bf k},\uparrow }c_{{\bf -k},\downarrow }\rangle
={\frac{\Delta
_{{\bf k}}}{{2\sqrt{\xi _{{\bf k}}^{2}+\Delta _{{\bf k}}^{2}}}}}\tanh {\frac{%
\sqrt{\xi _{{\bf k}}^{2}+\Delta _{{\bf k}}^{2}}}{{2T}},}
\end{equation}
and two coupled equations for the on-site and inter-site order
parameters are \index{order parameter}
\begin{equation}
\Delta _{0}=-{\frac{v_{0}}{{N}}}\sum_{{\bf k}}{\frac{\Delta _{{\bf k}}}{{2%
\sqrt{\xi _{{\bf k}}^{2}+\Delta _{{\bf k}}^{2}}}}\tanh \frac{\sqrt{\xi _{%
{\bf k}}^{2}+\Delta _{{\bf k}}^{2}}}{{2T}},}
\end{equation}
\begin{equation}
\Delta _{1}=-{\frac{v_{1}}{{Nw}}}\sum_{{\bf k}}{\frac{\Delta _{{\bf
k}}(\xi _{{\bf k}}+\mu )}{{2\sqrt{\xi _{{\bf k}}^{2}+\Delta _{{\bf
k}}^{2}}}}\tanh \frac{\sqrt{\xi _{{\bf k}}^{2}+\Delta _{{\bf
k}}^{2}}}{{2T}}.}
\end{equation}
These equations are equivalent to a single BCS equation \index{BCS equation} for $\Delta _{{\bf k}%
}=\Delta (\xi _{{\bf k}})$, but with the half polaron bandwidth $w$
cutting the integral, rather than the Debye temperature,
\begin{equation}
\Delta (\xi )=\int_{-w-\mu }^{w-\mu }d\eta N_{p}(\eta )V(\xi ,\eta ){\frac{%
\Delta (\eta )}{{2\sqrt{\eta ^{2}+\Delta (\eta )^{2}}}}\tanh \frac{\sqrt{%
\eta ^{2}+\Delta (\eta )^{2}}}{{2T}}.}
\end{equation}
Here $V(\xi ,\eta )=-v_{0}-zv_{1}{(\xi +\mu )(\eta +\mu )/w}^{2}$.

The critical temperature $T_{c}$ of the polaronic superconductor is
determined by two linearised equations  in the limit $%
\Delta _{0,1}\rightarrow 0$,
\begin{equation}
\left[ 1+A\left( {\frac{v_{0}}{{zv_{1}}}}+{\frac{\mu ^{2}}{{w^{2}}}}\right) %
\right] \Delta -{\frac{B\mu }{{w}}}\Delta _{1}=0,
\end{equation}
\begin{equation}
-{\frac{A\mu }{{w}}}\Delta +\left( 1+B\right) \Delta _{1}=0,
\end{equation}
where $\Delta =\Delta _{0}-\Delta _{1}{\mu /w,}$ and
\[
A={\frac{zv_{1}}{{2w}}}\int_{-w-\mu }^{w-\mu }d\eta {\frac{\tanh
{\frac{\eta }{{2T_{c}}}}}{\eta },}
\]
\[
B={\frac{zv_{1}}{{2w}}}\int_{-w-\mu }^{w-\mu }d\eta {\frac{\eta \tanh {\frac{%
\eta }{{2T_{c}}}}}{{w^{2}}}}.
\]
These equations are applied only if the polaron-polaron coupling is weak, $%
|v_{0,1}|<w$. A nontrivial solution is found at
\begin{equation}
T_{c}\approx 1.14w\sqrt{1-{\frac{\mu ^{2}}{{w^{2}}}}}\exp \left( {\frac{2w}{{%
v_{0}+zv_{1}{\mu ^{2}/}w^{2}}}}\right) ,
\end{equation}
if ${v_{0}+zv_{1}{\mu ^{2}/w}}^{2}<0,$ so that superconductivity
exists even in the case of the on-site repulsion, $v_{0}>0,$ if this
repulsion is less than the total intersite attraction, $z|v_{1}|$.
There is a nontrivial dependence of $T_{c}$ on doping. With a
constant density of states in the polaron band, the Fermi energy
$\epsilon _{F}\approx \mu $ is expressed via the number of polarons
per atom $n$ as
\begin{equation}
\mu =w(n-1),
\end{equation}
so that
\begin{equation}
T_{c}\simeq 1.14w\sqrt{n(2-n)}\exp \left( {\frac{2w}{{v_{0}+zv_{1}[n-1]^{2}}}%
}\right) ,
\end{equation}
which has two maxima as a function of $n$ separated by a deep
minimum in the half-filled band ($n=1$), where the nearest-neighbour
contributions to pairing are canceled.

\section{Mobile small bipolarons \index{bipolaron!small}}

The attractive energy of two small polarons is generally larger than
the polaron bandwidth, $\lambda -\mu _{c}\gg Z^{\prime }.$ When this
condition is satisfied, small bipolarons are not overlapped.
Consideration of particular lattice structures shows that small
bipolarons are mobile even when the electron-phonon coupling is
strong and the bipolaron binding energy is large \cite{ale5,jim}.
 Here we encounter a novel electronic state of matter, a
charged Bose liquid, qualitatively different from the normal
Fermi-liquid and  the BCS superfluid. The Bose-liquid is stable
because bipolarons repel each other (see below).

\subsection{On-site bipolarons and bipolaronic Hamiltonian
\index{bipolaron!on-site}}

The small parameter, $Z^{\prime }/(\lambda -\mu _{c})\ll 1,$ allows
for a consistent treatment of bipolaronic systems
\cite{ale0,aleran}. Under this condition the hopping term in the
transformed Hamiltonian $\tilde{H}$ is a small perturbation of the
ground state of immobile bipolarons and free phonons,
\begin{equation}
\tilde{H}=H_{0}+H_{pert},
\end{equation}
where
\begin{equation}
H_{0}={\frac{1}{{2}}}\sum_{i,j}v_{ij}c_{i}^{\dagger }c_{j}^{\dagger
}c_{j}c_{i}+\sum_{{\bf q,\nu }}\omega _{{\bf q\nu }}[d_{{\bf q\nu }%
}^{\dagger }d_{{\bf q\nu }}+1/2]
\end{equation}
and
\begin{equation}
H_{pert}=\sum_{i,j}\hat{\sigma}_{ij}c_{i}^{\dagger }c_{j}
\end{equation}
Let us first discuss the dynamics of {\it onsite }\ bipolarons,
which are the ground state of the system with the Holstein
non-dispersive e-ph interaction \cite{pwa,aleran}. The onsite
bipolaron is formed if
\begin{equation}
2E_{p}>U,
\end{equation}
where $U$ is the onsite Coulomb correlation energy (the  Hubbard $U$%
). The intersite polaron-polaron interaction \index{polaron!-polaron
interaction} is just the Coulomb repulsion since the phonon mediated
attraction between two polarons on different sites is zero in the
Holstein model. Two or more onsite bipolarons as well as three or
more polarons cannot occupy the same site because of the Pauli
exclusion principle. Hence, bipolarons repel single polarons and
each other. Their binding energy, $\Delta =2E_{p}-U,$
\index{bipolaron!binding energy} is larger than the polaron
half-bandwidth, $\Delta \gg w,$ so that there are no unbound
polarons in the ground state. $H_{pert}$, Eq.(97), destroys
bipolarons in the first order. Hence it has no diagonal matrix
elements. Then the bipolaron dynamics, including superconductivity,
is described by the use of a new canonical transformation $\exp
(S_{2})$ \cite{aleran}, which eliminates the first order of
$H_{pert}$,
\begin{equation}
(S_{2})_{fp}=\sum_{i,j}{\frac{\langle
f|\hat{\sigma}_{ij}c_{i}^{\dagger }c_{j}|p\rangle }{{E_{f}-E_{p}}}}.
\end{equation}
Here $E_{f,p}$ and $|f\rangle ,|p\rangle $ are the energy levels and
the
eigenstates of $H_{0}$. Neglecting the terms of the order higher than $%
(w/\Delta )^{2}$ we obtain
\begin{equation}
(H_{b})_{ff^{\prime }}\equiv \left(
e^{S_{2}}\tilde{H}e^{-S_{2}}\right) _{ff^{\prime }},
\end{equation}
\begin{eqnarray*}
(H_{b})_{ff^{\prime }} &\approx &(H_{0})_{ff^{\prime }}-{\frac{1}{{2}}}%
\sum_{\nu }\sum_{i\neq i^{\prime },j\neq j^{\prime }}\langle f|\hat{\sigma}%
_{ii^{\prime }}c_{i}^{\dagger }c_{i^{\prime }}|p\rangle \langle p|\hat{\sigma%
}_{jj^{\prime }}c_{j}^{\dagger }c_{j^{\prime }}|f^{\prime }\rangle \times \\
&&\left( {\frac{1}{{E_{p}-E_{f^{\prime }}}}}+{\frac{1}{{E_{p}-E_{f}}}}%
\right) .
\end{eqnarray*}
$S_{2}$ couples a localised onsite bipolaron and a state of two
unbound polarons on different sites. The expression (100) determines
the matrix elements of the transformed {\it \ bipolaronic}
Hamiltonian $H_{b}$ in the subspace $|f\rangle ,|f^{\prime }\rangle
$ with no single (unbound) polarons. On the other hand, the
intermediate {\it bra} $\langle p|$ and {\it ket} $|p\rangle $ refer
to configurations involving two unpaired polarons and any number of
phonons. Hence we have
\begin{equation}
E_{p}-E_{f}=\Delta +\sum_{{\bf q,\nu }}\omega _{{\bf q\nu }}\left( n_{{\bf %
q\nu }}^{p}-n_{{\bf q\nu }}^{f}\right) ,
\end{equation}
where $n_{{\bf q\nu }}^{f,p}$ are phonon occupation numbers $%
(0,1,2,3...\infty )$. This equation is an explicit definition of the
bipolaron binding energy $\Delta $ which takes into account the
residual inter-site repulsion between bipolarons and between two
unpaired polarons. The lowest eigenstates of $H_{b}$ are in the
subspace, which has only doubly occupied $c_{{\bf m}s}^{\dagger
}c_{{\bf m}s^{\prime }}^{\dagger }|0\rangle $ or empty $|0\rangle $
sites. On-site bipolaron tunnelling is a two-step transition. It
takes place via a single polaron tunneling to a neighbouring site.
The subsequent tunnelling of its "partner" to the same site restores
the initial energy state of the system. There are no $real$ phonons
emitted or absorbed because the (bi)polaron band is narrow. Hence we
can average $H_{b}$ with respect to phonons. Replacing the energy
denominators in the second term in Eq.(100) by the integrals with
respect to time,
\[
\frac{1}{{E_{p}-E_{f}}}=i\int_{0}^{\infty }dte^{i({E_{f}-Ep}+i\delta
)t},
\]
we obtain
\begin{eqnarray}
H_{b} &=&H_{0}-i\sum_{{\bf m\neq m}^{\prime },s}\sum_{{\bf n\neq
n}^{\prime
},s^{\prime }}T({\bf m-m}^{\prime })T({\bf n-n}^{\prime })\times \\
&&c_{{\bf m}s}^{\dagger }c_{{\bf m}^{\prime }s}c_{{\bf n}s^{\prime
}}^{\dagger }c_{{\bf n}^{\prime }s^{\prime }}\int_{0}^{\infty
}dte^{-i\Delta t}\Phi _{{\bf mm}^{\prime }}^{{\bf nn}^{\prime }}(t).
\nonumber
\end{eqnarray}
Here $\Phi _{{\bf mm}^{\prime }}^{{\bf nn}^{\prime }}(t)$ is a
multiphonon correlator,
\begin{equation}
\Phi _{{\bf mm}^{\prime }}^{{\bf nn}^{\prime }}(t)\equiv
\left\langle
\left\langle \hat{X}_{i}^{\dagger }(t)\hat{X}_{i^{\prime }}(t)\hat{X}%
_{j}^{\dagger }\hat{X}_{j^{\prime }}\right\rangle \right\rangle .
\end{equation}
 $%
\hat{X}_{i}^{\dagger }(t)$ and $\hat{X}_{i^{\prime }}(t)$ commute for any $%
\gamma ({\bf q,}\nu )=\gamma (-{\bf q,}\nu )$. $\hat{X}%
_{j}^{\dagger }$ and $\hat{X}_{j^{\prime }}$ commute as well, so
that we can write
\begin{eqnarray}
\hat{X}_{i}^{\dagger }(t)\hat{X}_{i^{\prime }}(t) &=&\prod_{{\bf q}%
}e^{[u_{i^{\prime }}({\bf q,}t)-u_{i}({\bf q,}t)]d_{{\bf q}}-H.c.]}, \\
\hat{X}_{j}^{\dagger }\hat{X}_{j^{\prime }} &=&\prod_{{\bf q}%
}e^{[u_{j^{\prime }}({\bf q})-u_{j}({\bf q})]d_{{\bf q}}-H.c.]},
\end{eqnarray}
where the phonon branch index $\nu $ is dropped for transparency.
Applying twice the identity Eq.(52) yields
\begin{eqnarray}
\hat{X}_{i}^{\dagger }(t)\hat{X}_{i^{\prime }}(t)\hat{X}_{j}^{\dagger }\hat{X%
}_{j^{\prime }} &=&\prod_{{\bf q}}e^{\beta ^{\ast }d_{{\bf
q}}^{\dagger
}}e^{-\beta d_{{\bf q}}}e^{-|\beta |^{2}/2}\times \\
&&e^{[u_{i^{\prime }}({\bf q,}t)-u_{i}({\bf q,}t)][u_{j^{\prime }}^{\ast }(%
{\bf q})-u_{j}^{\ast }({\bf q})]/2-H.c.},  \nonumber
\end{eqnarray}
where
\[
\beta =u_{i}({\bf q,}t)-u_{i^{\prime }}({\bf q,}t)+u_{j}({\bf q}%
)-u_{j}^{\prime }({\bf q}).
\]
Finally using the average Eq.(54) we find
\begin{eqnarray}
\Phi _{{\bf mm}^{\prime }}^{{\bf nn}^{\prime }}(t) &=&e^{-g^{2}({\bf m-m}%
^{\prime })}e^{-g^{2}({\bf n-n}^{\prime })}\times \\
&&\exp \left\{ \frac{1}{2N}\sum_{{\bf q,\nu }}|\gamma ({\bf q,}\nu )|^{2}F_{%
{\bf q}}({\bf m,m}^{\prime },{\bf n,n}^{\prime })\frac{\cosh \left[ \omega _{%
{\bf q\nu }}\left( \frac{1}{2T}-it\right) \right] }{\sinh \left[ \frac{%
\omega _{{\bf q\nu }}}{2T}\right] }\right\} ,  \nonumber
\end{eqnarray}
where
\begin{eqnarray}
F_{{\bf q}}({\bf m,m}^{\prime },{\bf n,n}^{\prime }) &=&\cos [{\bf q\cdot (n}%
^{\prime }-{\bf m})]+\cos [{\bf q\cdot (n}-{\bf m}^{\prime })]- \\
&&\cos [{\bf q\cdot (n}^{\prime }-{\bf m}^{\prime })]-\cos [{\bf q\cdot (n}-%
{\bf m})].  \nonumber
\end{eqnarray}

Taking into account that there are only bipolarons in the subspace, where $%
H_{b}$ \index{bipolaronic!Hamiltonian} operates, we finally rewrite
the Hamiltonian in terms of the creation $b_{{\bf m}}^{\dagger
}=c_{{\bf m}\uparrow }^{\dagger }c_{{\bf m}\downarrow
}^{\dagger }$ and annihilation $b_{{\bf m}}=c_{{\bf m}\downarrow }c_{{\bf m}%
\uparrow }$ operators of singlet pairs as
\begin{eqnarray}
H_{b} &=&-\sum_{{\bf m}}\left[ \Delta +{\frac{1}{{2}}}\sum_{{\bf m^{\prime }}%
}v^{(2)}({\bf m-m}^{\prime })\right] n_{{\bf m}}+ \\
&&\sum_{{\bf m\neq m^{\prime }}}\left[ t({\bf m-m}^{\prime })b_{{\bf m}%
}^{\dagger }b_{{\bf m^{\prime }}}+{\frac{1}{{2}}}\bar{v}({\bf
m-m}^{\prime })n_{{\bf m}}n_{{\bf m^{\prime }}}\right] .  \nonumber
\end{eqnarray}
There are no triplet pairs in the Holstein model, because the Pauli
exclusion principle does not allow two electrons with the same spin
to occupy the same site. Here $n_{{\bf m}}=b_{{\bf m}}^{\dagger
}b_{{\bf m}}$ is the bipolaron site-occupation operator,
\begin{equation}
\bar{v}({\bf m-m}^{\prime })=4v({\bf m-m}^{\prime })+v^{(2)}({\bf m-m}%
^{\prime }),
\end{equation}
is the bipolaron-bipolaron interaction \index{bipolaron!-bipolaron
interaction} including the direct polaron-polaron interaction
$v({\bf m-m}^{\prime })$ and a second order in $T({\bf m})$
repulsive correlation
\begin{equation}
v^{(2)}({\bf m-m}^{\prime })=2i\int_{0}^{\infty }dte^{-i\Delta t}\Phi _{{\bf %
mm^{\prime }}}^{{\bf m^{\prime }m}}(t).
\end{equation}
This additional repulsion appears because a virtual hop of one of
two polarons of the pair is forbidden, if the neighbouring site is
occupied by
another pair. The bipolaron transfer integral is of the second order in $T(%
{\bf m})$
\begin{equation}
t({\bf m-m}^{\prime })=-2iT^{2}({\bf m-m}^{\prime })\int_{0}^{\infty
}dte^{-i\Delta t}\Phi _{{\bf mm^{\prime }}}^{{\bf mm^{\prime }}}(t).
\end{equation}
The {\it bipolaronic} Hamiltonian, Eq.(109) describes the low-energy
physics of strongly coupled electrons and phonons. Using the
explicit form of the multiphonon correlator, Eq.(107), we obtain for
dispersionless phonons at $T\ll \omega _{0}$,
\begin{eqnarray*}
\Phi _{{\bf mm^{\prime }}}^{{\bf mm}^{\prime }}(t) &=&e^{-2g^{2}({\bf %
m-m^{\prime })}}\exp \left[ -2g^{2}({\bf m-m^{\prime })}e^{-i\omega _{0}t}%
\right] , \\
\Phi _{{\bf mm^{\prime }}}^{{\bf m}^{\prime }{\bf m}}(t) &=&e^{-2g^{2}({\bf %
m-m^{\prime })}}\exp \left[ 2g^{2}({\bf m-m^{\prime })}e^{-i\omega _{0}t}%
\right].
\end{eqnarray*}
 Expanding the time-dependent exponents in the Fourier series and
calculating the integrals in Eqs.(112) and (111) yield
\cite{alekab0}
\begin{equation}
t({\bf m)}=-{\frac{2T^{2}({\bf m})}{{\Delta }}}e^{-2g^{2}({\bf m)}%
}\sum_{l=0}^{\infty }\frac{[-2g^{2}({\bf m)}]^{l}}{l{!(1+l\omega
_{0}/\Delta )}}
\end{equation}
and
\begin{equation}
v^{(2)}({\bf m})={\frac{2T^{2}({\bf m})}{{\Delta }}}e^{-2g^{2}({\bf m)}%
}\sum_{l=0}^{\infty }\frac{[2g^{2}({\bf m)}]^{l}}{l{!(1+l\omega _{0}/\Delta )%
}}.
\end{equation}
When $\Delta \ll \omega _{0},$ we can keep the first term only with
$l=0$ in the bipolaron hopping integral, Eq.(113). In this case the
bipolaron half-bandwidth $zt({\bf a)}$ is of the order of
$2w^{2}/(z\Delta )$. However, if the bipolaron binding energy is
large, $\Delta \gg \omega _{0},$
the bipolaron bandwidth dramatically decreases proportionally to $%
e^{-4g^{2}}$ in the limit $\Delta \rightarrow \infty $. However,
this limit is not realistic because $\Delta
=2E_{p}-V_{c}<2g^{2}\omega _{0}$. In a more realistic regime,
$\omega _{0}<\Delta <2g^{2}\omega _{0}$, Eq.(113) yields
\begin{equation}
t({\bf m)}\approx {\frac{2\sqrt{2\pi }T^{2}({\bf m})}{\sqrt{\omega
_{0}\Delta }}}\exp \left[ -2g^{2}-{\frac{\Delta }{{\omega
}_{0}}}\left( 1+\ln \frac{{2g^{2}({\bf m)}\omega _{0}}}{\Delta
}\right) \right] .
\end{equation}
On the contrary, the bipolaron-bipolaron repulsion, Eq.(114) has no
small exponent in the limit $\Delta \rightarrow \infty $,
$v^{(2)}\propto
D^{2}/\Delta .$ Together with the direct Coulomb repulsion the second order $%
v^{(2)}$ ensures stability of the bipolaronic liquid against
clustering.

The high temperature behavior of the bipolaron bandwidth
\index{bipolaron!bandwidth} is just the opposite to that of the
small polaron bandwidth. While the polaron band collapses with
increasing temperature \cite{fir}, the bipolaron band becomes wider
\cite{bry2},
\begin{equation}
t({\bf m)}\propto \frac{1}{\sqrt{T}}\exp \left[ -{\frac{E_{p}+\Delta }{{2T}}}%
\right]
\end{equation}
for $T>\omega _{0}$.

\subsection{Superlight intersite bipolarons in the Fr\"ohlich-Coulomb model (FCM) \index{bipolaron!superlight}}

Any realistic theory of doped ionic insulators must include both the
long-range Coulomb repulsion between carriers and the strong
long-range electron-phonon interaction. From a theoretical
standpoint, the inclusion of the long-range Coulomb repulsion is
critical in ensuring that the carriers would not form clusters.
Indeed, in order to form stable bipolarons, the e-ph interaction has
to be strong enough to overcome the Coulomb repulsion at short
distances. Since the realistic e-ph interaction is long-ranged,
there is a potential possibility for clustering. The inclusion of
the Coulomb repulsion $V_{c}$ makes the clusters unstable. More
precisely, there is a certain window of $V_{c}/E_{p}$ inside which
the clusters are unstable but mobile bipolarons form nonetheless. In
this parameter window bipolarons repel each other and propagate in a
narrow band.

Let us consider a generic "Fr\"{o}hlich-Coulomb" Hamiltonian,
\index{Fr\"ohlich-Coulomb model} which explicitly includes the
infinite-range Coulomb and electron-phonon interactions, in a
particular lattice structure \cite{alekor2}. The implicitly present
infinite Hubbard $U$ prohibits double occupancy and removes the need
to distinguish the fermionic spin, as soon as we are interested in
the charge excitations alone. Introducing spinless fermion operators
$c_{{\bf n}}$ and phonon operators $d_{{\bf m}\nu }$, the
Hamiltonian is written as
\begin{eqnarray}
H &=&\sum_{{\bf n\neq n^{\prime }}}T({\bf n-n^{\prime }})c_{{\bf n}%
}^{\dagger }c_{{\bf n^{\prime }}}+{1\over{2}}\sum_{{\bf n\neq n^{\prime }}}V_{c}({\bf %
n-n^{\prime }})c_{{\bf n}}^{\dagger }c_{{\bf n}}c_{{\bf n^{\prime }}%
}^{\dagger }c_{{\bf n^{\prime }}}+ \\
&&\omega _{0}\sum_{{\bf n\neq m,}\nu }g_{\nu }({\bf m-n})({\bf
e}_{\nu }\cdot {\bf e}_{{\bf m-n}})c_{{\bf n}}^{\dagger }c_{{\bf
n}}(d_{{\bf m}\nu
}^{\dagger }+d_{{\bf m}\nu })+  \nonumber \\
&&\omega _{0}\sum_{{\bf m},\nu }\left( d_{{\bf m}\nu }^{\dagger }d_{{\bf m}%
\nu }+\frac{1}{2}\right).  \nonumber
\end{eqnarray}
The e-ph term is written in the real space representation (section
2), which is more convenient in working with complex lattices.

In general, the many-body model Eq.(117) is of considerable
complexity. However, we are interested in the non/near adiabatic
limit of the strong e-ph interaction. In this case, the kinetic
energy is a perturbation and the model can be grossly simplified
using the Lang-Firsov canonical transformation  in the Wannier
representation for electrons and phonons,
\[
S=\sum_{{\bf m\neq n,}\nu }g_{\nu }({\bf m-n})({\bf e}_{\nu }\cdot {\bf e}_{%
{\bf m-n}})c_{{\bf n}}^{\dagger }c_{{\bf n}}(d_{{\bf m}\nu }^{\dagger }-d_{%
{\bf m}\nu }).
\]
The transformed Hamiltonian is
\begin{eqnarray}
\tilde{H} &=&e^{-S}He^{S}=\sum_{{\bf n\neq n^{\prime }}}\hat{\sigma}_{{\bf %
nn^{\prime }}}c_{{\bf n}}^{\dagger }c_{{\bf n^{\prime }}}+\omega _{0}\sum_{%
{\bf m}\alpha }\left( d_{{\bf m}\nu }^{\dagger }d_{{\bf m}\nu }+\frac{1}{2}%
\right) + \\
&&\sum_{{\bf n\neq n^{\prime }}}v({\bf n-n^{\prime }})c_{{\bf
n}}^{\dagger
}c_{{\bf n}}c_{{\bf n^{\prime }}}^{\dagger }c_{{\bf n^{\prime }}}-E_{p}\sum_{%
{\bf n}}c_{{\bf n}}^{\dagger }c_{{\bf n}}.  \nonumber
\end{eqnarray}
The last term describes the energy gained by polarons due to e-ph
interaction. $E_{p}$ is the familiar polaron level shift,
\begin{equation}
E_{p}=\omega \sum_{{\bf m}\nu }g_{\nu }^{2}({\bf m-n})({\bf e}_{\nu
}\cdot {\bf e}_{{\bf m-n}})^{2},
\end{equation}
which is independent of ${\bf n}$. The third term on the right-hand
side in Eq.(130) is the polaron-polaron interaction:
\begin{equation}
v({\bf n-n^{\prime }})=V_{c}({\bf n-n^{\prime }})-V_{ph}({\bf n-n^{\prime }}%
),
\end{equation}
where
\begin{eqnarray*}
V_{ph}({\bf n-n^{\prime }}) &=&2\omega _{0}\sum_{{\bf m,}\nu }g_{\nu }({\bf %
m-n})g_{\nu }({\bf m-n^{\prime }})\times \\
&&({\bf e}_{\nu }\cdot {\bf e}_{{\bf m-n}})({\bf e}_{\nu }\cdot {\bf e}_{%
{\bf m-n^{\prime }}}).
\end{eqnarray*}
The phonon-induced interaction $V_{ph}$ is due to displacements of
common
ions by two electrons. The transformed hopping operator $\hat{\sigma%
}_{{\bf nn^{\prime }}}$ in the first term in Eq.(118) is given by
\begin{eqnarray}
\hat{\sigma}_{{\bf nn^{\prime }}} &=&T({\bf n-n^{\prime }})\exp \left[ \sum_{%
{\bf m,}\nu }\left[ g_{\nu }({\bf m-n})({\bf e}_{\nu }\cdot {\bf e}_{{\bf m-n%
}})\right. \right. \\
&&-\left. \left. g_{\nu }({\bf m-n^{\prime }})({\bf e}_{\nu }\cdot {\bf e}_{%
{\bf m-n^{\prime }}})\right] (d_{{\bf m}\alpha }^{\dagger }-d_{{\bf
m}\alpha })\right] .  \nonumber
\end{eqnarray}
This term perturbation at large $\lambda $. Here we consider a
particular lattice structure (ladder), where bipolarons tunnell
already in the first order in $T({\bf n})$, so that
$\hat{\sigma}_{{\bf nn^{\prime }}}$ can be averaged over phonons.
When $T \ll \omega _{0}$ the result is
\begin{equation}
t({\bf n-n^{\prime }})\equiv \left\langle \left\langle \hat{\sigma}_{{\bf %
nn^{\prime }}}\right\rangle \right\rangle _{ph}=T({\bf n-n^{\prime
}})\exp [-g^{2}({\bf n-n^{\prime }})],
\end{equation}
\begin{eqnarray*}
g^{2}({\bf n-n^{\prime }}) &=&\sum_{{\bf m},\nu }g_{\nu }({\bf m-n})({\bf e}%
_{\nu }\cdot {\bf e}_{{\bf m-n}})\times \\
&&\left[ g_{\nu }({\bf m-n})({\bf e}_{\nu }\cdot {\bf e}_{{\bf
m-n}})-g_{\nu
}({\bf m-n^{\prime }})({\bf e}_{\nu }\cdot {\bf e}_{{\bf m-n^{\prime }}})%
\right] .
\end{eqnarray*}
The mass renormalization exponent can be expressed via $E_{p}$ and
$V_{ph}$ as
\begin{equation}
g^{2}({\bf n-n^{\prime }})=\frac{1}{\omega _{0}}\left[ E_{p}-\frac{1}{2}%
V_{ph}({\bf n-n^{\prime }})\right] .
\end{equation}
Now phonons are "integrated out" and the polaronic Hamiltonian is
given by
\begin{equation}
H_{p}=H_{0}+H_{pert},
\end{equation}
\[
H_{0}=-E_{p}\sum_{{\bf n}}c_{{\bf n}}^{\dagger }c_{{\bf
n}}+{1\over{2}}\sum_{{\bf n\neq
n^{\prime }}}v({\bf n-n^{\prime }})c_{{\bf n}}^{\dagger }c_{{\bf n}}c_{{\bf %
n^{\prime }}}^{\dagger }c_{{\bf n^{\prime }}},
\]
\[
H_{pert}=\sum_{{\bf n\neq n^{\prime }}}t({\bf n-n^{\prime }})c_{{\bf n}%
}^{\dagger }c_{{\bf n^{\prime }}}.
\]
When $V_{ph}$ exceeds $V_{c}$ the full interaction becomes negative
and polarons form pairs. The real space representation allows us to
elaborate more physics behind the lattice sums in Eq.(119) and
Eq.(120) \cite{alekor2}. When a carrier (electron or hole) acts on
an ion with a force ${\bf f}$, it displaces the ion by some vector
${\bf x}={\bf f}/k$. Here $k$ is the ion's
force constant. The total energy of the carrier-ion pair is $-{\bf f}%
^{2}/(2k)$. This is precisely the summand in Eq.(119) expressed via
dimensionless coupling constants. Now consider two carriers
interacting with
the {\em same} ion,  Fig.4a. The ion displacement is ${\bf x}=({\bf f}%
_{1}+{\bf f}_{2})/k$ and the energy is $-{\bf f}_{1}^{2}/(2k)-{\bf f}%
_{2}^{2}/(2k)-({\bf f}_{1}\cdot {\bf f}_{2})/k$. Here the last term
should be interpreted as an ion-mediated interaction between the two
carriers. It depends on the scalar product of ${\bf f}_{1}$ and
${\bf f}_{2}$ and consequently on the relative positions of the
carriers with respect to the ion. If the ion is an isotropic
harmonic oscillator, as we assume here, then the following simple
rule applies. If the angle $\phi $ between ${\bf f}_{1}$ and ${\bf
f}_{2}$ is less than $\pi /2$ the polaron-polaron interaction will
be attractive, if otherwise it will be repulsive. In general, some
ions will generate attraction, and some repulsion between polarons,
Fig. 4b.

\begin{figure}[tbp]
\begin{center}
\includegraphics[angle=-0,width=0.65\textwidth]{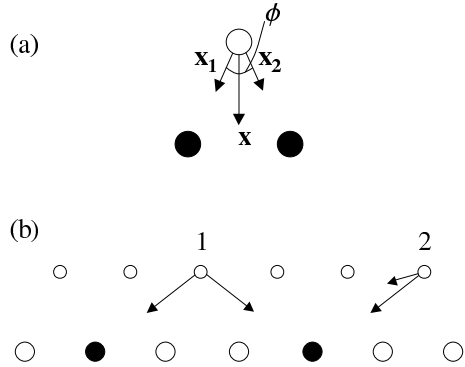}
\end{center}
\caption{Mechanism of the polaron-polaron interaction. (a) Together,
 two polarons (solid circles) deform the lattice more
effectively than separately. An effective attraction occurs when the
angle $\phi$ between ${\bf x}_{1}$ and ${\bf x}_{2}$ is less than
$\pi /2$. (b) A mixed situation: ion 1 results in repulsion between
two polarons while ion 2 results in attraction.}
\end{figure}

The overall sign and magnitude of the interaction is given by the
lattice sum in Eq.(120), the evaluation of which is elementary. One
should also note that according to Eq.(123) an attractive e-ph
interaction reduces the polaron mass (and consequently the bipolaron
mass), while repulsive e-ph interaction enhances the mass. Thus, the
long-range nature of the e-ph interaction serves a double purpose.
Firstly, it generates an additional inter-polaron attraction because
the distant ions have small angle $\phi $. This additional
attraction helps to overcome the direct Coulomb repulsion between
 polarons. And secondly, the Fr\"{o}hlich interaction makes the
bipolarons lighter.

The many-particle ground state of $H_{0}$ depends on the sign of the
polaron-polaron interaction, \index{polaron!-polaron interaction}
the carrier density, and the lattice geometry. Here we consider the
zig-zag ladder, Fig.5a, assuming that all sites are isotropic
two-dimensional harmonic oscillators. For simplicity, we also
adopt the nearest-neighbour approximation for both interactions, $g_{\nu }(%
{\bf l})\equiv g$, $V_{c}({\bf n})\equiv V_{c}$, and for the hopping
integrals, $T({\bf m})=T_{NN}$ for $l=n=m=a$, and zero otherwise.
Hereafter we set the lattice period $a=1$. There are four nearest
neighbours in the ladder, $z=4$. Then the {\it one-particle}
polaronic Hamiltonian takes the form
\begin{eqnarray}
H_{p} &=&-E_{p}\sum_{n}(c_{n}^{\dagger }c_{n}+p_{n}^{\dagger }p_{n})+ \\
&&\sum_{n}[t^{\prime }(c_{n+1}^{\dagger }c_{n}+p_{n+1}^{\dagger
}p_{n})+t(p_{n}^{\dagger }c_{n}+p_{n-1}^{\dagger }c_{n})+H.c.],
\nonumber
\end{eqnarray}
where $c_{n}$ and $p_{n}$ are polaron annihilation operators on the
lower and upper legs of the ladder, respectively, Fig.5b. Using
Eqs.(119), (120) and (123) we find
\begin{eqnarray}
E_{p} &=&4g^{2}\omega _{0}, \\
t^{\prime } &=&T_{NN}\exp \left( -\frac{7E_{p}}{8\omega _{0}}\right)
,
\nonumber \\
t &=&T_{NN}\exp \left( -\frac{3E_{p}}{4\omega _{0}}\right) .
\nonumber
\end{eqnarray}
The Fourier transform of Eq.(125) into momentum space yields
\begin{eqnarray}
H_{p} &=&\sum_{k}(2t^{\prime }\cos k-E_{p})(c_{k}^{\dagger
}c_{k}+p_{k}^{\dagger }p_{k})+ \\
&&t\sum_{k}[(1+e^{ik})p_{k}^{\dagger }c_{k}+H.c.].  \nonumber
\end{eqnarray}
A linear transformation of $c_{k}$ and $p_{k}$ diagonalises the
Hamiltonian. There are two overlapping polaronic bands,
\begin{equation}
E_{1}(k)=-E_{p}+2t^{\prime }\cos (k)\pm 2t\cos (k/2)
\end{equation}
with the effective mass $m^{\ast }=2/|4t^{\prime }\pm t|$ near their
edges.

Let us now place two polarons on the ladder. The nearest neighbour
interaction is $v=V_{c}-E_{p}/2,$ if two polarons are on different
legs of the ladder, and $v=V_{c}-E_{p}/4,$ if both polarons are on
the same leg. The attractive interaction is provided via the
displacement of the lattice sites, which are the common nearest
neighbours to both polarons. There are two such nearest neighbours
for the intersite bipolaron of type $A$ or $B$, Fig.5c, but there is
only one common nearest neighbour for bipolaron $C$, Fig.5d. When
$V_{c}>E_{p}/2$, there are no bound states and the multi-polaron
system is a one-dimensional Luttinger liquid. However, when
$V_{c}<E_{p}/2$ and consequently $v<0$, the two polarons are bound
into an inter-site bipolaron \index{bipolaron!inter-site} of types
$A$ or $B$.

It is quite remarkable that the bipolaron tunnelling
\index{bipolaron!tunnelling} in the ladder already appears in the
first order with respect to a single-electron tunnelling. This case
is different from both onsite bipolarons discussed above, and from
intersite chain bipolarons of Ref. \cite{bon}, where the bipolaron
tunnelling was of the second order in $T(a)$. Indeed, the lowest
energy degenerate configurations $A$ and $B$ are degenerate. They
are coupled by $H_{pert}.$ Neglecting all higher-energy
configurations, we can project the Hamiltonian onto the subspace
containing $A$, $B$, and empty sites.

\begin{figure}[tbp]
\begin{center}
\includegraphics[angle=-0,width=0.65\textwidth]{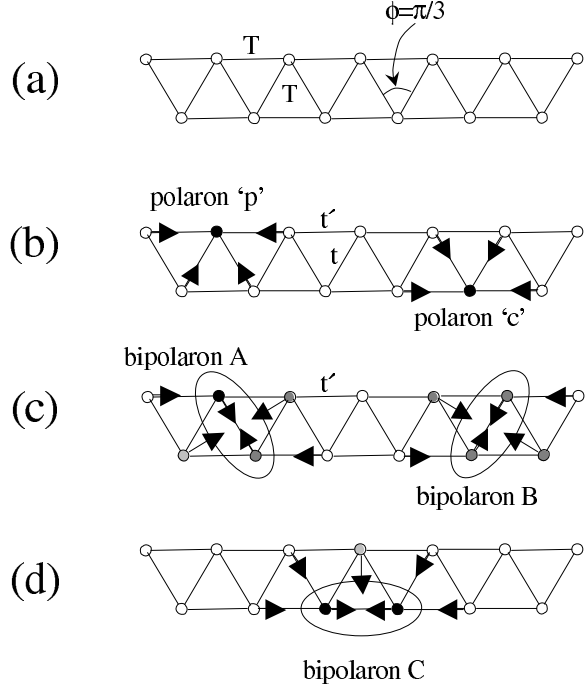}
\end{center}
\caption{One-dimensional zig-zag ladder. (a) Initial ladder with the
bare hopping amplitude $T(a)$. (b) Two types of polarons with their
respective deformations. (c) Two degenerate bipolaron configurations
A and B. (d) A different bipolaron configuration, C, whose energy is
higher than that of A and B.}
\end{figure}
The result of such a projection is the bipolaronic Hamiltonian
\begin{equation}
H_{b}=\left( V_{c}-\frac{5}{2}E_{p}\right) \sum_{n}[A_{n}^{\dagger
}A_{n}+B_{n}^{\dagger }B_{n}]-t^{\prime }\sum_{n}[B_{n}^{\dagger
}A_{n}+B_{n-1}^{\dagger }A_{n}+H.c.],
\end{equation}
where $A_{n}=c_{n}p_{n}$ and $B_{n}=p_{n}c_{n+1}$ are intersite
bipolaron annihilation operators, and the bipolaron-bipolaron
interaction is dropped (see below). The Fourier transform of
Eq.(129) yields two {\it bipolaron} bands,
\begin{equation}
E_{2}(k)=V_{c}-{\frac{5}{{2}}}E_{p}\pm 2t^{\prime }\cos (k/2).
\end{equation}
with a combined width $4|t^{\prime }|$. The bipolaron binding energy
in zero order with respect to $t,t^{\prime }$ is
\begin{equation}
\Delta \equiv 2E_{1}(0)-E_{2}(0)=\frac{E_{p}}{2}-V_{c}.
\end{equation}
The bipolaron mass near the bottom of the lowest band, $m^{\ast \ast
}=2/t^{\prime }$, is
\begin{equation}
m^{\ast \ast }=4m^{\ast }\left[ 1+0.25\exp \left( \frac{{E_{p}}}{8\omega _{0}%
}\right) \right] .
\end{equation}
The numerical coefficient $1/8$ in the exponent ensures that
$m^{\ast \ast }$ remains of the order of $m^{\ast }$ even at large
$E_{p}$. This fact combines with a weaker renormalization of
$m^{\ast }$ providing a {\em superlight} bipolaron.
\index{bipolaron!superlight}

In models with strong intersite attraction there is a possibility of
clusterisation. Similar to the two-particle case above, the lowest
energy of $n$ polarons placed on the nearest neighbours of the
ladder is found as
\begin{equation}
E_{n}=(2n-3)V_{c}-\frac{6n-1}{4}E_{p}
\end{equation}
for any $n\geq 3$. There are {\em no} resonating states for a
$n$-polaron configuration if $n\geq 3$. Therefore there is no
first-order kinetic energy
contribution to their energy. $E_{n}$ should be compared with the energy $%
E_{1}+(n-1)E_{2}/2$ of far separated $(n-1)/2$ bipolarons and a
single polaron for odd $n\geq 3$, or with the energy of far
separated $n$ bipolarons for even $n\geq 4$. ``Odd'' clusters are
stable if
\begin{equation}
V_{c}<\frac{n}{6n-10}E_{p},
\end{equation}
and ``even'' clusters are stable if
\begin{equation}
V_{c}<\frac{n-1}{6n-12}E_{p}.
\end{equation}
As a result we find that bipolarons repel each other and single polarons at $%
V_{c}>3E_{p}/8$. If $V_{c}$ is less than $3E_{p}/8$ then immobile
bound clusters of three and more polarons could form. One should
notice that at distances much larger than the lattice constant the
polaron-polaron interaction is always repulsive, and the formation
of infinite clusters, stripes or strings is impossible (see also
\cite{kabanov}). Combining the condition of bipolaron formation and
that of the instability of larger clusters we obtain a window of
parameters
\begin{equation}
\frac{3}{8}E_{p}<V_{c}<\frac{1}{2}E_{p},
\end{equation}
where the ladder is a bipolaronic conductor. Outside the window the
ladder is either charge-segregated into finite-size clusters (small
$V_{c}$), or it is a liquid of repulsive polarons (large $V_{c}$).

There is strong experimental evidence for superlight intersite
bipolarons  in cuprate superconductors (see below), where they form
in-plane oxygen - apex oxygen pairs (so called apex bopolarons)
and/or in-plane oxygen-oxygen pairs \cite{ale5,cat,alekor2}. While
the long-range Fr\"ohlich interaction \index{Fr\"ohlich!interaction}
combined with Coulomb repulsion might cause clustering of polarons
into finite-size quasi-metallic mesoscopic textures,
\index{mesoscopic textures} the analytical \cite{alekabstring} and
QMC \cite{merkabmic} studies of mesoscopic textures with lattice
deformations and Coulomb repulsion show that pairs  dominate over
phase separation \index{phase separation} since bipolarons
effectively repel each other (see also \cite{kabanov}.

\section{Bipolaronic superconductivity \index{superconductivity!bipolaronic}}

In the subspace with no single polarons, the Hamiltonian of
electrons strongly-coupled with phonons is reduced to the
bipolaronic Hamiltonian \index{bipolaronic!Hamiltonian} written in
terms of creation, $b_{{\bf m}}^{\dagger }=c_{{\bf m}\uparrow
}^{\dagger }c_{{\bf m}\downarrow }^{\dagger }$ and annihilation, $b_{{\bf m}%
} $, bipolaron operators as
\begin{equation}
H_{b}=\sum_{{\bf m\neq m^{\prime }}}\left[ t({\bf m-m}^{\prime })b_{{\bf m}%
}^{\dagger }b_{{\bf m^{\prime }}}+{\frac{1}{{2}}}\bar{v}({\bf
m-m}^{\prime })n_{{\bf m}}n_{{\bf m^{\prime }}}\right] ,
\end{equation}
where $\bar{v}({\bf m-m}^{\prime })$ is the bipolaron-bipolaron
interaction, $n_{{\bf m}}=b_{{\bf m}}^{\dagger }b_{{\bf m}}$, and
the position of the middle of the bipolaron band
\index{bipolaron!band} is taken as zero. There are additional spin
quantum numbers $S=0,1;S_{z}=0,\pm 1,$ which should be added to the
definition of $b_{{\bf m}}$ in the case of intersite bipolarons,
which tunnel via the one-particle hopping. This "crab-like"
tunnelling, Fig.5, results in a bipolaron bandwidth of the same
order as the polaron one. Keeping this in mind we can apply $H_{b}$,
Eq.(137) to both on-site and/or inter-site bipolarons,
\index{bipolaron!inter-site} \index{bipolaron!on-site} and even to
more extended non-overlapping pairs, implying that the site index
${\bf m}$ is the position of the centre of mass of a pair.

Bipolarons are not perfect bosons. In the subspace of pairs and
empty sites their operators commute as
\begin{equation}
b_{{\bf m}}b_{{\bf m}}^{\dagger }+b_{{\bf m}}^{\dagger }b_{{\bf
m}}=1,
\end{equation}
\begin{equation}
b_{{\bf m}}b_{{\bf m^{\prime }}}^{\dagger }-b_{{\bf m^{\prime
}}}^{\dagger }b_{{\bf m}}=0
\end{equation}
for ${\bf m}\neq {\bf m^{\prime }}$. This makes useful the
pseudospin analogy \cite{aleran},
\begin{equation}
b_{{\bf m}}^{\dagger }=S_{{\bf m}}^{x}-iS_{{\bf m}}^{y}
\end{equation}
and
\begin{equation}
b_{{\bf m}}^{\dagger }b_{{\bf m}}={\frac{1}{{2}}}-S_{{\bf m}}^{z}
\end{equation}
with the pseudospin $1/2$ operators $S^{x,y,z}={\frac{1}{{2}}}\tau _{1,2,3}$%
. $S_{{\bf m}}^{z}=1/2$ corresponds to an empty site ${\bf m}$ and $S_{{\bf m%
}}^{z}=-1/2$ to a site occupied by the bipolaron. Spin operators
preserve the bosonic nature of bipolarons, when they are on
different sites, and their fermionic internal structure. Replacing
bipolarons by spin operators we transform the bipolaronic
Hamiltonian into the anisotropic Heisenberg Hamiltonian,
\index{Heisenberg Hamiltonian}
\begin{equation}
H_{b}=\sum_{{\bf m\neq m^{\prime }}}\left[ {\frac{1}{{2}}}\bar{v}_{{\bf %
mm^{\prime }}}S_{{\bf m}}^{z}S_{{\bf m^{\prime }}}^{z}+t_{{\bf mm^{\prime }}%
}(S_{{\bf m}}^{x}S_{{\bf m^{\prime }}}^{x}+S_{{\bf m}}^{y}S_{{\bf m^{\prime }%
}}^{y})\right] .
\end{equation}
This Hamiltonian has been investigated in detail as a relevant form
for magnetism and also for quantum solids like a lattice model of
$^{4}He$. However, while in those cases the magnetic field is an
independent thermodynamic variable, in our case the total
``magnetization'' is fixed,
\begin{equation}
{\frac{1}{{N}}}\sum_{{\bf m}}\langle \langle S_{{\bf m}}^{z}\rangle \rangle =%
{\frac{1}{{2}}}-n_{b},
\end{equation}
if the bipolaron density $n_{b}$ is conserved. Spin $1/2$ Heisenberg
Hamiltonian, Eq.(142) cannot be solved analytically. Complicated
commutation rules for bipolaron operators make the
problem hard, but not in the limit of low atomic density of bipolarons, $%
n_{b}\ll 1$ (for a complete phase diagram of bipolarons on a lattice
see Refs. \cite{aleran,alerob}). In this limit we can reduce the
problem to a charged Bose gas on a lattice \cite{sam}. Let us
transform the bipolaronic
Hamiltonian to a representation containing only the Bose operators $a_{{\bf m%
}}$ and $a_{{\bf m}}^{\dagger }$ defined as
\begin{equation}
b_{{\bf m}}=\sum_{k=0}^{\infty }\beta _{k}(a_{{\bf m}}^{\dagger })^{k}a_{%
{\bf m}}^{k+1},
\end{equation}
\begin{equation}
b_{{\bf m}}^{\dagger }=\sum_{k=0}^{\infty }\beta _{k}(a_{{\bf
m}}^{\dagger })^{k+1}a_{{\bf m}}^{k},
\end{equation}
where
\begin{equation}
a_{{\bf m}}a_{{\bf m^{\prime }}}^{\dagger }-a_{{\bf m^{\prime
}}}^{\dagger }a_{{\bf m}}=\delta _{{\bf m,m^{\prime }}}.
\end{equation}
The first few coefficients $\beta _{k}$ are found by substituting
Eqs.(144) and (145) into Eqs.(138) and (139),
\begin{equation}
\beta _{0}=1,\beta _{1}=-1,\beta
_{2}={\frac{1}{{2}}}+{\frac{\sqrt{3}}{{6}}.}
\end{equation}
We also introduce bipolaron and boson $\Psi $-operators as
\begin{equation}
\Phi ({\bf r})={\frac{1}{\sqrt{N}}}\sum_{{\bf m}}\delta ({\bf r-m})b_{{\bf m}%
},
\end{equation}
\begin{equation}
\Psi ({\bf r})={\frac{1}{\sqrt{N}}}\sum_{{\bf m}}\delta ({\bf r-m})a_{{\bf m}%
}.
\end{equation}
The transformation of the field operators takes the form
\begin{equation}
\Phi ({\bf r})=\left[ 1-{\frac{\Psi ^{\dagger }({\bf r})\Psi ({\bf r})}{{N}}}%
+{\frac{(1/2+\sqrt{3}/6)\Psi ^{\dagger }({\bf r})\Psi ^{\dagger }({\bf r})\Psi ({\bf r})\Psi ({\bf r})%
}{{N^{2}}}}+...\right] \Psi ({\bf r}).
\end{equation}
Then we write the bipolaronic Hamiltonian as
\begin{equation}
H_{b}=\int d{\bf r}\int d{\bf r^{\prime }}\Psi ^{\dagger }({\bf r})t({\bf %
r-r^{\prime }})\Psi ({\bf r^{\prime }})+H_{d}+H_{h}+H^{(3)},
\end{equation}
where
\begin{equation}
H_{d}={\frac{1}{{2}}}\int d{\bf r}\int d{\bf r^{\prime }}\bar{v}({\bf %
r-r^{\prime }})\Psi ^{\dagger }({\bf r})\Psi ^{\dagger }({\bf r^{\prime }}%
)\Psi ({\bf r}^{\prime })\Psi ({\bf r}),
\end{equation}
is the dynamic part,
\begin{eqnarray}
H_{k} &=&{\frac{2}{{N}}}\int d{\bf r}\int d{\bf r^{\prime }}t({\bf %
r-r^{\prime }})\times \\
&&\left[ \Psi ^{\dagger }({\bf r})\Psi ^{\dagger }({\bf r^{\prime }})\Psi (%
{\bf r^{\prime }})\Psi ({\bf r^{\prime }})+\Psi ^{\dagger }({\bf
r})\Psi ^{\dagger }({\bf r})\Psi ({\bf r})\Psi ({\bf r^{\prime
}})\right] . \nonumber
\end{eqnarray}
is the kinematic (hard-core) part due to the "imperfect" commutation
rules, and $H^{(3)}$ includes three- and higher-body collisions.
Here
\[
t({\bf r-r^{\prime }})=\sum_{{\bf k}}\epsilon _{{\bf k}}^{\ast \ast }e^{i%
{\bf k\cdot (r-r^{\prime })}},
\]
\[
\bar{v}({\bf r-r^{\prime }})={\frac{1}{{N}}}\sum_{{\bf k}}\bar{v}_{{\bf k}%
}e^{i{\bf k\cdot (r-r^{\prime })}},
\]
$\bar{v}_{{\bf k}}=\sum_{{\bf m}\neq 0}\bar{v}({\bf m)}\exp (i{\bf k\cdot m}%
) $ is the Fourier component of the dynamic interaction and
\begin{equation}
\epsilon _{{\bf k}}^{\ast \ast }=\sum_{{\bf m\neq }0}t({\bf m})\,\exp (-i%
{\bf k\cdot m})
\end{equation}
is the bipolaron band dispersion. \index{bipolaron!band!dispersion}
$H^{(3)}$ contains powers of the field operator higher than four. In
the dilute limit, $n_{b}\ll 1,$ only two-particle interactions are
essential which include the short-range kinematic and direct
density-density repulsions. Because $\bar{v}$ already has the short
range part $v^{(2)},$ Eq.(113), the kinematic contribution can be
included in the definition of $\bar{v}$. As a result $H_{b}$ is
reduced to the Hamiltonian of interacting hard-core charged bosons
tunnelling in the narrow band.

To describe electrodynamics of bipolarons \index{bipolaron!electrodynamics} we introduce the vector potential $%
{\bf A}({\bf r})$ using the so-called Peierls substitution
\cite{pei},
\[
t({\bf m-m^{\prime })}\rightarrow t({\bf m-m^{\prime })}e^{i2e{\bf A}({\bf m}%
)\cdot ({\bf m-m^{\prime }})},
\]
which is a fair approximation when the magnetic field is weak
compared with the atomic field, $eHa^{2}<<1$. It has the following
form,
\begin{equation}
t({\bf r-r^{\prime })}\rightarrow t({\bf r,r^{\prime }})=\sum_{{\bf k}%
}\epsilon _{{\bf k}-2e{\bf A}}^{\ast \ast }e^{i{\bf k\cdot
(r-r^{\prime })}}
\end{equation}
in real space. If the magnetic field is weak, we can expand $\epsilon _{{\bf %
k}}^{\ast \ast }$ in the vicinity of ${\bf k}=0$ to obtain
\begin{equation}
t({\bf r,r^{\prime }})\approx -{\frac{\left[ {\bf \nabla +}2ie{\bf A}({\bf r}%
)\right] ^{2}}{{2m^{\ast \ast }}}}\delta ({\bf r-r^{\prime }}),
\end{equation}
where
\begin{equation}
\frac{1}{{m^{\ast \ast }}}=\left( {\frac{d^{2}\epsilon _{{\bf
k}}^{\ast \ast }}{{d}k{^{2}}}}\right) _{k\rightarrow 0}
\end{equation}
is the inverse bipolaron mass. Here we assume a parabolic dispersion
near the bottom of the band, $\epsilon _{{\bf k}}^{\ast \ast }\sim
k^{2}$, so that
\begin{eqnarray}
H_{b} &\approx &-\int d{\bf r}\Psi ^{\dagger }({\bf r})\left\{ {\frac{[{\bf %
\nabla +}2ie{\bf A}({\bf r})]^{2}}{{2m^{\ast \ast }}}}+\mu \right\} \Psi (%
{\bf r})+ \\
&&{\frac{1}{{2}}}\int d{\bf r}d{\bf r^{\prime }}\bar{v}({\bf r-r^{\prime }}%
)\Psi ^{\dagger }({\bf r})\Psi ^{\dagger }({\bf r^{\prime }})\Psi ({\bf r}%
)\Psi ({\bf r^{\prime }}),  \nonumber
\end{eqnarray}
where we add the {\it bipolaron} chemical potential $\mu $. We note
that
the bipolaron-bipolaron interaction is the Coulomb repulsion, $\bar{v}({\bf r%
})\sim 1/(\epsilon _{0}r)$ at large distances, and the hard-core
repulsion is not important in the dilute limit. The Hamiltonian
Eq.(159) describes the charged Bose gas  with the effective boson
mass $m^{\ast \ast }$ and charge $2e$, which is a superconductor
\cite{shaf}.

\section{Bipolaronic superconductivity in  cuprates \index{bipolaronic!superconductivity!in cuprates}}

 The fact that  Helium-4 and its
isotope Helium-3 are well known Bose and Fermi superfluids,
respectively,  with very different superfluid transition
temperatures
($T_{c}=2.17K$ in $^4He$ and $%
T_{c}=0.0026K$ in $^3He$)  already kindles the view that
high-temperature superconductivity might derive from preformed
real-space charged bosons rather than in the BCS state with  Cooper
pairs, which are correlations  in momentum space. The possibility of
real-space pairing of carriers in cuprates, as opposed to Cooper
pairing, has been the subject of much discussion. Some authors
dismissed any real-space pairs even in underdoped cuprates, where a
low density of carriers appears to favor individual pairing rather
than Cooper pairing. But on the other hand real-space pairing is
strongly supported by our strong-coupling extension  of the BCS
theory since the e-ph interaction is very strong in cuprates. Also
on the experimental side a growing  number of other independent
observations point to the possibility that high-$T_{c}$
superconductors may not be conventional Bardeen-Cooper-Schrieffer
(BCS) superconductors, \index{superconductor!BCS} but rather derive
from the Bose-Einstein condensation (BEC) \index{Bose-Einstein
condensation} of real-space superlight small bipolarons. There is
strong evidence for real-space pairing and the three-dimensional BEC
in cuprates from unusual upper critical fields \cite{aleH}
\index{critical field} and isotope effects \cite{aleiso}
\index{isotope effect} predicted by us
 and the $\lambda$-like electronic specific heat \cite{alekablia},
parameter-free fitting of experimental $T_c$ with BEC $T_{c}$
\cite{alekabfit}, normal state pseudogaps \cite{alegap,kabmic}
\index{pseudogap} and anisotropy \cite{alekabmot}, and more recently
from normal state diamagnetism \cite{aledia}, \index{diamagnetism}
the Hall-Lorenz numbers \cite{leeale,alelor}, the normal state
Nernst effect \cite{alezav,alecom}, and the giant proximity effect
(GPE) \cite{alepro}.  Here I briefly discuss a few of these
remarkable observations (for more details see Ref. \cite{alebook}
and Part IV). \index{Nernst effect} \index{proximity effect}
\index{Hall-Lorenz number}

\subsection{Upper critical field,  the Hall-Lorenz number, and  isotope effects}
Magnetotransport \cite{zavkabale} and thermal magnetotransport
\cite{zha,leeale} data strongly support preformed pairs in cuprates.
In particular,
 many high magnetic field studies revealed a non-BCS upward
curvature of the upper critical field, $H_{c2}(T)$ and its
non-linear temperature dependence in the vicinity of $\ T_{c}$  in a
number of cuprates as well as in a few other unconventional
superconductors, Fig.6. If unconventional superconductors are in the
"bosonic" limit of preformed real-space pairs, such unusual critical
fields are  expected in accordance with the  theoretical prediction
for the Bose-Einstein condensation of charged bosons in an external
magnetic field \cite{aleH}.
\begin{figure}[tbp]
\begin{center}
\includegraphics[angle=-0,width=0.80\textwidth]{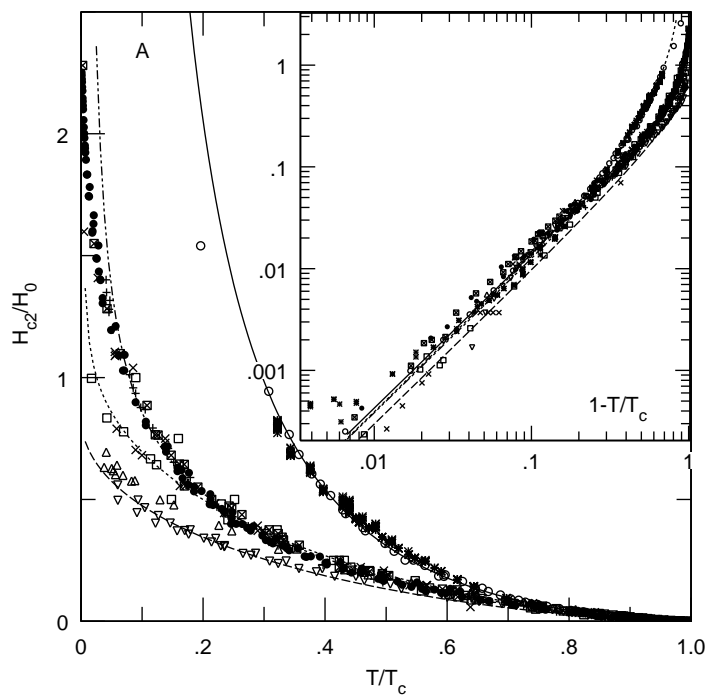} \vskip -0.5mm
\end{center}
\caption{\small{Resistive upper critical field \cite{zavkabale}
(determined at 50\% of the transition) of cuprates, spin-ladders and
organic superconductors scaled according to the Bose-Einsten
condensation field of charged bosons \cite{aleH},
$H_{c2}(T)\propto\left[b(1-t)/t+1-t^{1/2}\right] ^{3/2}$ with
$t=T/T_c$.  The parameter $b$ is proportional to the number of
delocalised bosons at zero temperature, $b$ is 1 (solid line), 0.02
(dashed-dotted line), 0.0012 (dotted line), and 0 (dashed line). The
inset shows a universal scaling of the same data near $T_{c}$ on the
logarithmic scale. Symbols correspond to $Tl-2201(\bullet) $,
$La_{1.85}Sr_{0.15}CuO_{4} (\triangle)$, $Bi-2201 (\times)$,
$Bi-2212 (\ast)$, $YBa_{2}Cu_3O_{6+x}(\circ)$,
$La_{2-x}Ce_{x}CuO_{4-y}$ (squares), $Sr_2Ca_{12}Cu_{24}O_{41} (+)$,
and Bechgaard salt organic superconductor ($\nabla)$ }.}
\end{figure}

Notwithstanding, some "direct" evidence for the existence of a
charge $2e$ Bose liquid in the normal state of cuprates is highly
desirable.   Alexandrov and Mott \cite {NEV} discussed the thermal
conductivity $\kappa $; the contribution from the carriers given by
the Wiedemann-Franz ratio depends strongly on the elementary charge
as $\sim (e^{\ast })^{-2}$ and should be significantly suppressed in
the case of $e^{\ast }=2e$ compared with the Fermi-liquid
contribution. As a result, the Lorenz number, $L=\left(
e/k_{B}\right)
^{2}\kappa _{e}/(T\sigma )$ differs significantly from the Sommerfeld value $%
L_{e}=\pi ^{2}/3$ of the standard Fermi-liquid theory, if carriers
are double-charged bosons. Here $\kappa _{e}$, $\sigma $, and $e$
are the electronic thermal conductivity, the electrical
conductivity, and the elementary charge, respectively. Ref.
\cite{NEV} predicted a rather low Lorenz number for bipolarons,
$L=6L_{e}/(4\pi ^{2})\approx 0.15L_{e}$, due to the double charge of
carriers, and also due to their nearly classical distribution
function above $T_{c}$.

 The extraction of the electron thermal conductivity \index{thermal
 conductivity}
has proven difficult since both the electron term, $\kappa _{e}$ and
the phonon term, $\kappa _{ph}$ are comparable to each other in the
cuprates. A new way to determine the Lorenz number has been realized
by Zhang et al. \cite{zha}, based on the thermal Hall conductivity.
The thermal Hall effect allowed for an efficient way to separate the
phonon heat current
even when it is dominant. As a result, the "Hall" Lorenz number, $%
L_{H}=\left( e/k_{B}\right) ^{2}\kappa _{xy}/(T\sigma _{xy})$, has
been
directly measured in $YBa_{2}Cu_{3}O_{6.95}$ because transverse thermal $%
\kappa _{xy}$ and electrical $\sigma _{xy}$ conductivities involve
only the electrons. Remarkably, the measured value of $L_{H}$ just
above $T_{c}$ is about the same as predicted by the bipolaron model,
$L_{H}\approx 0.15L_{e}. $ The experimental $L_{H}$ showed a strong
temperature dependence, which violates the Wiedemann-Franz law. This
experimental observation has been accounted for by taking into
account thermally excited polarons and also triplet pairs  in the
bipolaron model \cite{leeale}, Fig.7.
\begin{figure}[tbsp]
\begin{center}
\includegraphics[angle=-00,width=0.80\textwidth]{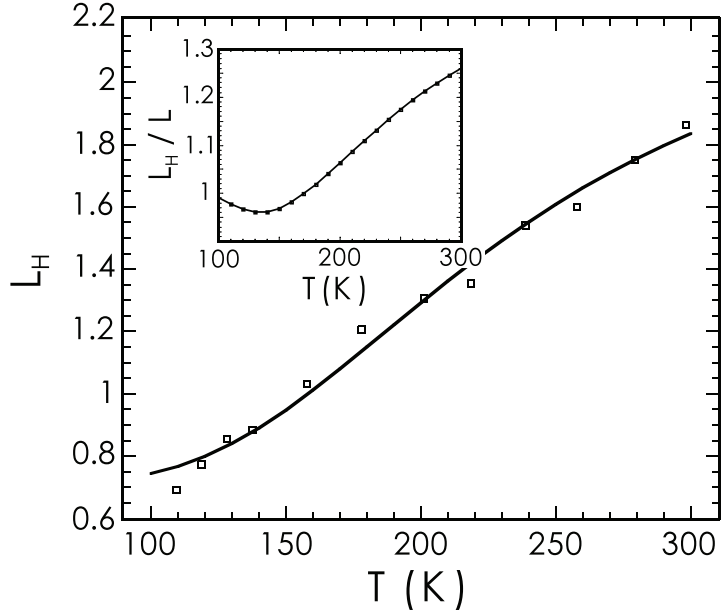}
\end{center}
\caption{The Hall Lorenz number $L_H$ \cite{leeale} of charged
bosons fits the experiment in YBa$_2$Cu$_3$O$_{6.95}$ \cite{zha}.
The pseudogap is taken as 675 K. The inset gives the ratio of the
Hall Lorenz number to the Lorenz number in the model.}
\end{figure}

Another compelling evidence for (bi)polaronic carries in novel
superconductors was provided by the discovery of  substantial
isotope effects on $T_c$ and  on the carrier mass
\cite{zhao,zhaochapter}. The advances in the fabrication of the
isotope substituted samples made it possible to measure a sizable
isotope effect , $\alpha =-d\ln T_{c}/d\ln M$ in many high-$T_{c}$
oxides. This led to a general conclusion that phonons are relevant
for high $T_{c}$. Moreover the isotope effect in cuprates was found
to be quite different from the BCS prediction, $\alpha =0.5$ (or
less). Several compounds showed $\alpha > 0.5$ , and sometimes
negative values of $\alpha $ were observed.

Essential features of the isotope effect, \index{isotope effect} in particular large values in low $%
T_{c}$ cuprates, an overal trend to lower value as $T_{c}$
increases, and a small or even negative $\alpha $ in some high
$T_{c}$ cuprates can be understood in the framework of the bipolaron
theory \cite{aleiso}. With increasing ion mass the
bipolaron mass increases and the Bose-Einstein condensation temperature $%
T_{c} \propto 1/m^{**}$ decreases in the bipolaronic superconductor
(section 8). On the contrary in polaronic superconductors (section
6) an increase of the ion mass leads to a band narrowing enhancing
 the polaron density of states and increasing $T_{c}$.
Hence the isotope exponent of $T_{c}$ can distinguish the BCS like
polaronic superconductivity with $\alpha < 0$ , and the
Bose-Einstein condensation of small bipolarons with $\alpha > 0$.
Moreover, underdoped cuprates, which are certainly  in the BEC
regime, could have $\alpha
>0.5,$  as observed.

The isotope effect on $T_{c}$ is linked with the  isotope effect on
the carrier mass, $\alpha_{m^*}$, as \cite{aleiso}
\begin{equation}
\alpha=-d\ln T_c/d\ln M=\alpha_{m^*}[1-Z/(\lambda-\mu_c)],
\end{equation}
where $\alpha_{m^*}= d \ln m^*/d \ln M$ and $Z=m/m^* \ll 1$. In
ordinary metals, where the Migdal approximation is believed to be
valid, the renormalized effective mass of electrons is independent
of the ion mass $M$ because the electron-phonon interaction constant
$\lambda$ does not depend on $M$. However, when the e-ph interaction
is sufficiently strong, the electrons form polarons dressed by
lattice distortions, with an effective mass $m^{\ast} = m \exp
(\gamma E_p/\omega)$. While $E_p$ in the above expression does not
depend on the ion mass, the phonon frequency does.  As a result,
there is a large isotope effect on the carrier mass in polaronic
conductors, $\alpha_{m^*} = (1/2)\ln (m^*/m)$ \cite{aleiso}, in
contrast to the zero isotope effect in ordinary metals. Such an
effect was observed in cuprates  in the London penetration depth
$\lambda_H$ of isotope-substituted samples \cite{zhao}. The carrier
density is unchanged with the isotope substitution of $O^{16}$ by
$O^{18}$, so that the isotope effect on $\lambda_H$ measures
directly the isotope effect on the carrier mass. In particular, the
carrier mass isotope exponent $\alpha
_{m^*}$ was found as large as $\alpha _{m^*}=0.8$ in $%
La_{1.895}Sr_{0.105}CuO_{4}$.

More recent high resolution angle resolved photoemission
spectroscopy \cite{lan0} \index{angle resolved photoemission}
provided another compelling evidence for a strong e-ph interaction
in  cuprates. It revealed a fine phonon structure in the electron
self-energy of the underdoped La$_{2-x}$Sr$_x$CuO$_4$ samples.
Remarkably,  an isotope effect on the electron spectral function in
Bi-2212 \cite{lan2} has been discovered. \index{isotope effect}
These experiments together with a number of earlier optical
\cite{mic1,ita,tal,tim,dev2,dev0} and
 neutron-scattering \cite{ega} experimental and theoretical
 studies firmly established the
 strong
coupling of carries with  optical phonons in cuprates (see also Part
IV).

\subsection{Normal state diamagnetism: BEC versus phase
fluctuations}

 Above $T_{c}$ the charged
bipolaronic  Bose liquid is non-degenerate and below $T_{c}$ phase
coherence (ODLRO) of the preformed bosons sets in. The state above
$T_{c}$ is perfectly "normal" in the sense that the off-diagonal
order parameter (i.e. the Bogoliubov-Gor'kov anomalous average
\index{Bogoliubov averages} $\cal{F}(\mathbf{r,r^{\prime }})=\langle
\psi_{\downarrow }(\mathbf{{r})\psi _{\uparrow }({r^{\prime
}}\rangle}$) is zero above the resistive transition temperature
$T_{c}$ as in the BCS theory. Here $\psi_{\downarrow,\uparrow
}(\mathbf{r})$ annihilates electrons with spin $\downarrow,
\uparrow$ at point ${\bf r}$.

\begin{figure}
\begin{center}
\includegraphics[angle=-0,width=0.80\textwidth]{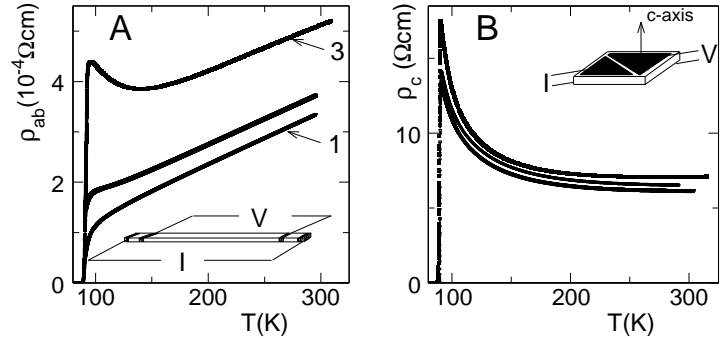}
\vskip -0.5mm \caption{In-plane (A) and out-of-plane (B) resistivity
of 3   single crystals of Bi$_2$Sr$_2$CaCu$_2$O$_8$ \cite{zavale}
showing no signature of phase fluctuations well above the resistive
transition. }
\end{center}
\end{figure}

However in contrast with the bipolaron and BCS theories a
significant fraction of research in the field of cuprate
superconductors suggests a so-called phase fluctuation scenario
\cite{kiv,xu,ong}, where $\cal{F}(\mathbf{r,r^{\prime }})$ remains
nonzero well above $T_{c}$. I believe that the phase fluctuation
scenario is impossible to reconcile with the extremely sharp
resistive transitions at $T_c$ in high-quality underdoped, optimally
doped and overdoped cuprates. For example, the in-plane and
out-of-plane resistivity of Bi-2212, where the anomalous Nernst
signal has been measured \cite{xu}, is perfectly "normal" above
$T_c$, Fig.8, showing only a few percent positive or negative
magnetoresistance \cite{zavale}, explained with bipolarons
\cite{zavalemos}.

Both in-plane \cite{buc,mac0,boz,lawrie,gan} and out-of-plane \cite
{alezavnev,hof2,zve} resistive transitions  of high-quality samples
 remain sharp in the magnetic field providing a
reliable determination of the genuine $H_{c2}(T)$.  The preformed
Cooper-pair (or phase fluctuation) model \cite{kiv}  is incompatible
with a great number of thermodynamic, magnetic, and kinetic
measurements, which show that only holes (density $x$), doped into a
parent insulator are carriers \emph{both} in  the normal and the
superconducting states of cuprates. The assumption \cite{kiv} that
the superfluid density $x$ is small compared with the normal-state
carrier density is also inconsistent with the theorem \cite{pop},
which proves that the number of supercarriers at $T=0$K  should be
the same as the number of normal-state carriers in any  clean
superfluid.

The normal state diamagnetism \index{diamagnetism!of cuprates} of
cuprates provides another clear evidence for BEC rather than for the
phase fluctuation scenario. A number of experiments (see, for
example, \cite{mac,junM,hof,nau,igu,ong} and references therein),
including torque magnetometries, showed enhanced diamagnetism above
$T_c$, which has been explained as the fluctuation diamagnetism in
quasi-2D superconducting cuprates (see, for example Ref.
\cite{hof}). The data taken at relatively low magnetic fields
(typically below 5 Tesla) revealed a crossing point in the
magnetization $M(T,B)$ of most anisotropic cuprates (e.g.
$Bi-2212$), or in $M(T,B)/B^{1/2}$ of less anisotropic $YBCO$
\cite{junM}. The dependence of magnetization (or $M/B^{1/2}$) on the
magnetic field has been shown to vanish at some characteristic
temperature below $T_c$. However the data taken in high magnetic
fields (up to 30 Tesla) have shown that the crossing point,
anticipated for low-dimensional superconductors and associated with
superconducting fluctuations, does not explicitly exist in magnetic
fields above 5 Tesla \cite{nau}.

Most surprisingly the torque magnetometery  \cite{mac,nau} uncovered
a diamagnetic signal somewhat above $T_c$ which increases in
magnitude with applied magnetic field.
\begin{figure}
\begin{center}
\includegraphics[angle=-90,width=0.75\textwidth]{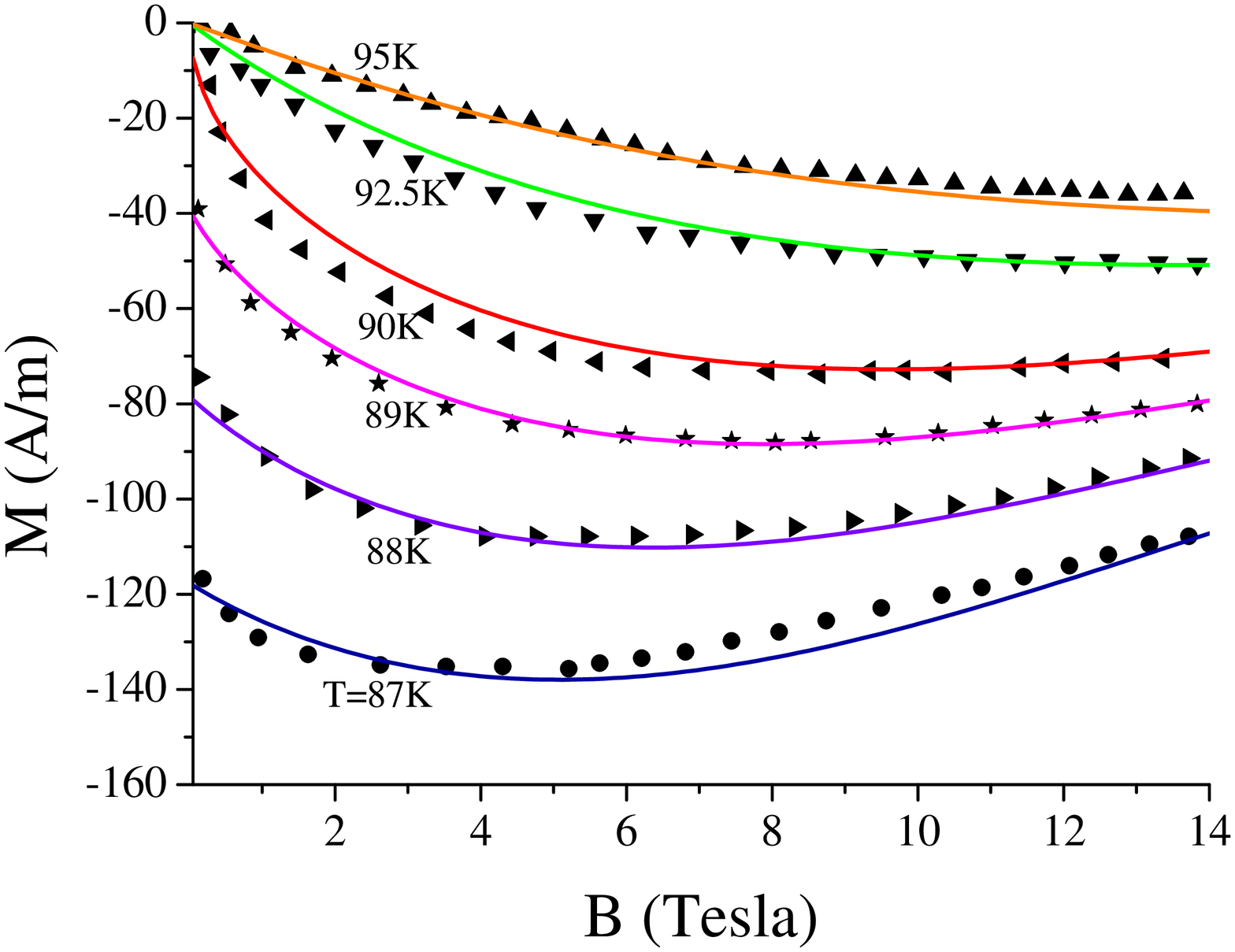}
\vskip -0.5mm \caption{Diamagnetism of optimally doped Bi-2212
(symbols)\cite{ong} compared with magnetization of CBG \cite{aledia}
near and above $T_c$ (lines).}
\end{center}
\end{figure}
It has been  linked with the Nernst signal and mobile vortexes   in
the  normal state of cuprates \cite{ong}. \index{Nernst effect}
However, apart from the inconsistences mentioned above, the vortex
scenario of the normal-state diamagnetism is internally
inconsistent. Accepting the vortex scenario and fitting  the
magnetization data in $Bi-2212$ with the conventional  logarithmic
field dependence \cite{ong}, one obtains surprisingly high upper
critical fields $H_{c2} > 120$ Tesla and a very large
Ginzburg-Landau parameter, $\kappa=\lambda/\xi
>450$  even at temperatures close to $T_c$. The in-plane
low-temperature magnetic field penetration depth is $\lambda=200$ nm
in optimally doped $Bi-2212$ (see, for example \cite{tallon}). Hence
the zero temperature coherence length $\xi$ turns out to be about
the lattice constant, $\xi=0.45$nm, or even smaller. Such a small
coherence length rules out the "preformed Cooper pairs"  \cite{kiv},
since the pairs are virtually not overlapped at any size of the
Fermi surface in $Bi-2212$ . Moreover the magnetic field dependence
of $M(T,B)$ at and above $T_c$ is entirely inconsistent  with what
one expects from a vortex liquid.  While $-M(B)$  decreases
logarithmically at temperatures well below $T_c$, the  experimental
curves \cite{mac,nau,ong} clearly show that   $-M(B)$  increases
with the field at and  above $T_c$ , just opposite to what one could
expect in the vortex liquid.  This significant departure from the
London liquid behavior clearly indicates that the vortex liquid does
not appear above the resistive phase transition \cite{mac}.
\index{vortex liquid}

Some time ago we  explained the anomalous diamagnetism in cuprates
as the Landau normal-state diamagnetism of preformed bosons
\cite{den}.  More recently the model has been extended   to high
magnetic fields taking into account the magnetic pair-breaking of
singlet bipolarons and the anisotropy of the energy spectrum
\cite{aledia}. When the
 magnetic field is applied perpendicular to the copper-oxygen
plains the quasi-2D bipolaron energy spectrum is quantized as
$E_\alpha= \omega(n+1/2) +2t_c [1-\cos(K_zd)]$, where $\alpha$
comprises $n=0,1,2,...$ and  in-plane $K_x$ and out-of-plane $K_z$
center-of-mass quasi-momenta, $\omega=2 eB/\sqrt{m^{\ast \ast}_x
m^{\ast \ast}_y}$, $t_c$ and $d$ are the hopping integral and the
lattice period perpendicular to the planes. We assume here that the
spectrum consists of two degenerate branches, so-called $"x"$ and
$"y"$ bipolarons  as in the case of apex intersite pairs \cite{ale5}
with anisotropic in-plane bipolaron masses $m^{\ast \ast}_x\equiv m$
and $m^{\ast \ast}_y\approx 4m$. Expanding the Bose-Einstein
distribution function in powers of $\exp[(\mu-E)/T]$ with the
negative chemical potential $\mu$ one can after summation over $n$
readily obtain
 the boson density
\begin{equation}
n_b={2eB\over{\pi  d}} \sum_{r=1}^{\infty} I_0(2t_c r/T) {\exp[
(\mu-\omega/2 -2t_c)r/T]\over{1-\exp(-\omega r/T)}},
\end{equation}
and the magnetization,
\begin{eqnarray}
&&M(T,B)=-n_b \mu_b+  {2eT\over{\pi  d}} \sum_{r=1}^{\infty}
I_0\left({2t_c r\over {T}}\right)\times \\
&&{\exp[ (\mu-\omega/2 -2t_c)r/T]\over{1-\exp(- \omega r/T)}}
\left({1\over{r}}-{\omega \exp(-\omega r/T)\over{k_BT[1-\exp(-\omega
r/T)]}}\right).\nonumber
\end{eqnarray}
Here $\mu_b= e/\sqrt{m^{\ast \ast}_xm^{\ast \ast}_y}$
 and $I_0(x)$ is the modified Bessel
function. At low temperatures $T \rightarrow 0$ Schafroth's result
\cite{shaf} is recovered, $M(0,B)= -n_b \mu_b$. The magnetization of
charged bosons is field-independent at low temperatures. At high
temperatures, $T \gg T_c$ the chemical potential has a large
magnitude, and we can keep only the terms with $r=1$ in
Eqs.(160,161) to obtain $M(T,B)=-n_b \mu_b \omega/(6T)$ at $T \gg
T_c\gg \omega$,
 which is the familiar  Landau  orbital diamagnetism  of nondegenerate
 carriers. Here $T_c$ is the  Bose-Einstein condensation
temperature $T_{c}= 3.31(n_{b}/2)^{2/3}/(m^{\ast \ast}_{x}m^{\ast
\ast}_{y}m^{\ast \ast}_{c})^{1/3}$, with $m_{c}=1/2|t_{c}|d^{2}$.

Comparing with experimental data one has to take into account a
temperature and field depletion of singlets due to their thermal
excitations into spin-split triplet states,
 $n_b(T,B)=n_c[1-\alpha \tau -(B/B^*)^2]$. Here
 $\alpha=3(2n_ct)^{-1}[J (e^{J/T_c}-1)^{-1}-T_c\ln(1-e^{-J/T_c})]$,
 $\mu_BB^*=(2T_cn_ct)^{1/2}
\sinh(J/2T_c)$, $\mu_B\approx 0.93 \times 10^{-23}$ Am$^2$ is the
 Bohr magneton, $n_c$ is the density of singlets at $T=T_c$  in zero
 field, $\tau=T/T_c-1$, $J$ is the singlet-triplet exchange energy,
 and $2t$ is the triplet bandwidth. As a result,  Eq.(161) fits remarkably well the  experimental curves in the critical
region of  optimally doped Bi-2212, Fig.9,   with $n_c \mu_b=
2100$A/m, $T_c=90$K,  $\alpha=0.62$ and $B^*=56$ Tesla, which
corresponds to the singlet-triplet exchange energy $J\approx 20$K.

On the other hand the experimental data, Fig.9,  contradict BCS and
the phase-fluctuation scenarios \cite{kiv,ong}.
\index{phase-fluctuation scenario}  Indeed, if we define a critical
exponent as $\delta=\ln B/\ln|M(T,B)|$ for $B\rightarrow 0$, the $T$
dependence of $\delta(T)$ in the charged Bose gas (CBG) is
dramatically different from the Berezinski-Kosterlitz-Thouless (BKT)
transition \index{Berezinski-Kosterlitz-Thouless transition}
critical exponents (as proposed in the phase fluctuation scenario),
but it is very close to the experimental \cite{ong} $\delta(T)$,
Fig.10.
\begin{figure}
\begin{center}
\includegraphics[angle=-90,width=0.75\textwidth]{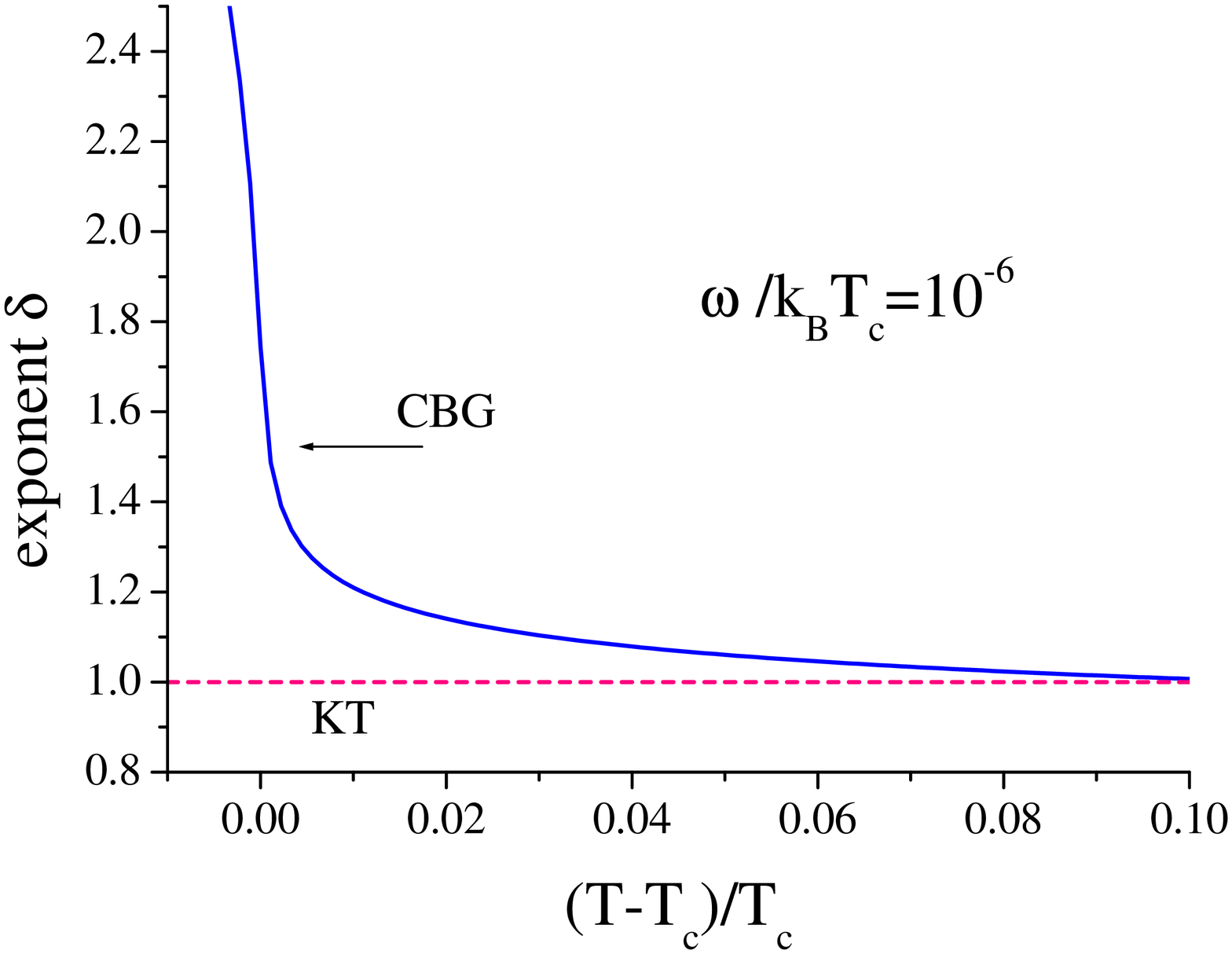}
\vskip -0.5mm \caption{Critical exponents of the low-field
magnetization in CBG and in BKT transition.}
\end{center}
\end{figure}

Also the large Nernst signal, allegedly supporting vortex liquid in
the normal state of cuprates \cite{ong}, has been  explained as the
normal state phenomenon owing to a partial localization of charge
carriers in a random potential inevitable in cuprates \cite{alezav}.
The coexistence of the large Nernst signal and the insulating-like
resistivity in slightly doped cuprates sharply disagrees with the
vortex scenario, but agrees remarkably well with our theory
\cite{alecom}.

\subsection{Giant proximity effect \index{proximity effect}}
Several groups reported that in the Josephson cuprate \emph{SNS}
junctions  supercurrent can run through normal \emph{N}-barriers as
thick as  100 nm in a strong conflict with the standard theoretical
picture, if the barrier is made from non-superconducting cuprates.
Using an advanced molecular beam epitaxy, Bozovic \emph{et al.}
\cite{bozp} proved that this giant proximity effect (GPE) is
intrinsic, rather than extrinsic caused by any inhomogeneity of the
barrier. Hence GPE defies the conventional explanation, which
predicts that the critical current should exponentially decay with
the characteristic length of about the coherence length, which is
$\xi \leq 1$ nm in the cuprates.

This effect can be broadly understood as the Bose-Einstein
condensate  tunnelling into
 a cuprate
\emph{semiconductor} \cite{alepro}. Indeed the chemical potential
$\mu$ remains in the charge-transfer gap of doped cuprates like
La$_{2-x}$Sr${_x}$CuO$_4$ \cite{boz0}  because of the bipolaron
formation. The condensate wave function, $\psi(Z)$,  is described by
the Gross-Pitaevskii (GP) equation.   In the superconducting region,
$Z<0$, near the $SN$ boundary, Fig.11, the equation is
\begin{equation}
{1\over{2m^{**}_c}}{d^2\psi(Z)\over{dZ^2}}=[V
|\psi(Z)|^2-\mu]\psi(Z),
\end{equation}
where $V$ is a short-range repulsion of bosons, and $m^{**}_c$ is
the boson mass along $Z$.  Deep inside the superconductor
$|\psi(Z)|^2=n_s$ and $\mu=Vn_s$ , where the condensate density
$n_s$ is about $x/2$, if the temperature is well below $T_c$  of the
superconducting electrode (the in-plane lattice constant $a$ and the
unit cell volume are
 taken as unity).

The normal barrier  at $Z
>0$ is an underdoped cuprate semiconductor above its
transition temperature, where the chemical potential $\mu$ lies
below the bosonic band by some energy $\epsilon$, Fig.11.   For
quasi-two dimensional bosons one readily obtains \cite{alebook}
\begin{equation}
\epsilon(T)=-T\ln(1-e^{-T_0/T}),
\end{equation}
where $T_0=\pi x'/m^{**}$, $m^{**}$ is the in-plane boson mass, and
$x' < x$ is the doping level of the barrier. Then the GP equation in
the barrier can be written as
\begin{equation}
{1\over{2m^{**}_c}}{d^2\psi(Z)\over{dZ^2}} =[V
|\psi(Z)|^2+\epsilon]\psi(Z).
\end{equation}
Introducing the bulk coherence length, $\xi= 1/(2m^{**}_c
n_sV)^{1/2}$ and dimensionless $f(z)=\psi(Z)/n_s^{1/2}$,
$\tilde{\mu}=\epsilon/n_sV$, and $z=Z/\xi$, one obtains
 for a real
$f(z)$
\begin{equation}
{d^2f\over{dz^2}} =f^3-f,
\end{equation}
if $z<0$, and
\begin{equation}
{d^2f\over{dz^2}}=f^3+\tilde{\mu}f,
\end{equation}
if
 $z>0$. These equations can be readily solved using  first
integrals of motion respecting the boundary conditions,
$f(-\infty)=1$, and $f(\infty)=0$,
\begin{equation}
{df\over{dz}}= -(1/2+f^4/2-f^2)^{1/2},
\end{equation}
and
\begin{equation}
{df\over{dz}}= -(\tilde{\mu}f^2+f^4/2)^{1/2},
\end{equation}
for $z<0$ and $z>0$, respectively. The solution in the
superconducting electrode is given by
\begin{equation}
f(z)=\tanh \left[-2^{-1/2}z+0.5
\ln{{2^{1/2}(1+\tilde{\mu})^{1/2}+1}\over{2^{1/2}(1+\tilde{\mu})^{1/2}-1}}\right].
\end{equation}
It decays  in the close vicinity of the barrier from 1 to
$f(0)=[2(1+\tilde{\mu})]^{-1/2}$ in the interval about the coherence
length $\xi$. On the other side  of the boundary, $z>0$, it is given
by
\begin{equation}
f(z)={(2\tilde{\mu})^{1/2}\over{\sinh\{z\tilde{\mu}^{1/2}+\ln[2(\tilde{\mu}(1+\tilde{\mu}))^{1/2}+(1+4\tilde{\mu}(1+\tilde{\mu}))^{1/2}]\}}}
.
\end{equation}

\begin{figure}
\begin{center}
\includegraphics[angle=-90,width=0.85\textwidth]{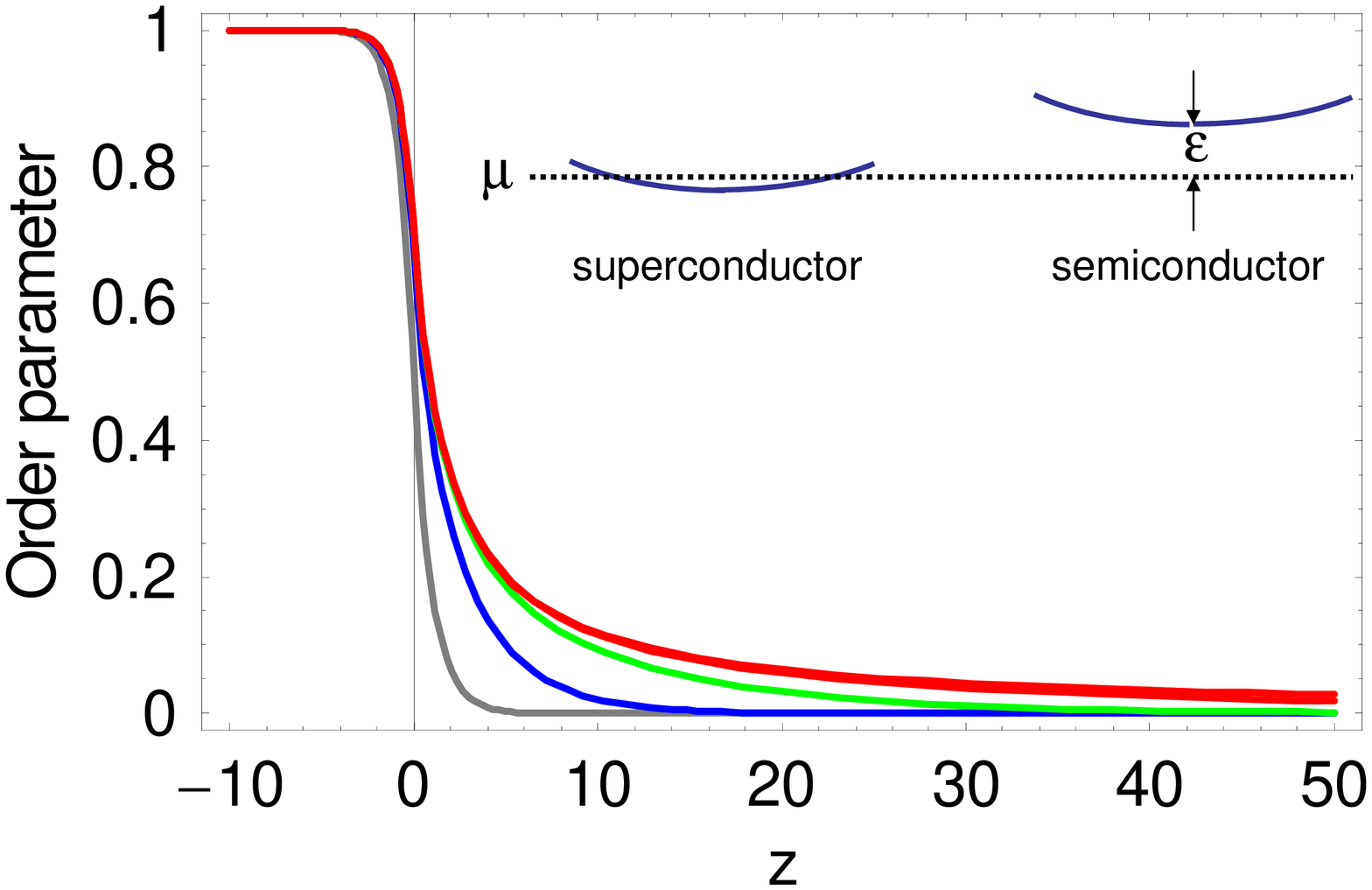}
\vskip -0.5mm \caption{BEC order parameter at the $SN$ boundary for
$\tilde{\mu}=1.0,0.1,0.01$ and $ \leq 0.001$ (upper curve).}
\end{center}
\end{figure}

Its profile is shown in Fig.11. Remarkably, the order parameter
\index{order parameter} penetrates into the normal layer up to the
length $Z^* \approx (\tilde{\mu})^{-1/2}\xi$, which could be larger
than $\xi$ by many orders of magnitude,  if $\tilde{\mu}$ is  small.
It is indeed the case, if the barrier layer is sufficiently doped.
For example, taking $x'=0.1$,   c-axis $m^{**}_c=2000 m_e$, in-plane
$m^{**}=10 m_e$ \cite{alebook}, $a=0.4$ nm, and $\xi=0.6$ nm, yields
 $T_0\approx 140$ K and $(\tilde{\mu})^{-1/2}\approx 5000$ at $T=10$K. Hence the
 order parameter could penetrate into the normal cuprate semiconductor
 up to more than a thousand coherence lengths as observed \cite{bozp}. If the thickness of the barrier $L$ is small compared with $Z^*$,
and $(\tilde{\mu})^{1/2}\ll 1$, the order parameter decays following
 the power law, rather than exponentially,
\begin{equation}
f(z)={\sqrt{2}\over{z+2}}.
\end{equation}
Hence, for $L \leq Z^*$, the critical current should also decay
following the power law \cite{alepro}. On the other hand, for an
\emph{undoped}
 barrier $\tilde{\mu}$ becomes
 larger than unity, $\tilde{\mu} \propto \ln(m^{**}T/\pi x')\rightarrow \infty$ for any finite temperature $T$  when $x' \rightarrow
 0$, and the current should exponentially decay with the characteristic length  smaller that $\xi$, as is experimentally observed as well \cite{boz0}.
As a result the bipolaron theory accounts for the giant and nil
proximity effects
 in slightly doped semiconducting and undoped insulating
cuprates, respectively. It predicts  the occurrence of a new length
scale, $\hbar/\sqrt{2m^{**}_c\epsilon (T)}$,  and explains the
temperature dependence of the critical current of $SNS$ junctions
\cite{alepro}.

\section{Conclusion}

Extending the BCS theory towards the strong interaction between
electrons and ion vibrations, a charged Bose gas \index{charged Bose
gas} of tightly bound small bipolarons was predicted by us
\cite{aleran} with a further prediction  that high $T_c$ should
exist in the crossover region of the e-ph interaction strength from
the BCS-like to bipolaronic superconductivity \cite{ale0}.

For very strong electron-phonon coupling, polarons become
self-trapped on a single lattice site. The energy of the resulting
small polaron is given as $-E_p=-\lambda zt$. Expanding about the
atomic limit in  hopping integrals $t$ (which is small compared to
$E_p$ in the small polaron regime, $\lambda>1$)
 the polaron mass is
computed as $m^{*}=m_0\exp(\gamma E_p/\omega_0)$ , where $\omega_0$
is the frequency of Einstein phonons, $m_0$ is the rigid band mass
on a cubic lattice, and $\gamma$ is a numerical constant. For the
Holstein model, which is purely site local, $\gamma=1$. Bipolarons
are on-site singlets in the Holstein model and their mass
$m_{H}^{**}$ appears only in the second order of $t$ \cite{aleran}
scaling as $m_{H}^{**}\propto (m^{*})^2$ for $\omega\gg\Delta$ , and
as $m_{H}^{**}\propto(m^*)^{4}$ in a more realistic regime
$\omega\ll\Delta$ (section 7). Here $\Delta=2E_p-U$ is the bipolaron
binding energy, and $U$ is the on-site (Hubbard) repulsion.
\index{Hubbard!repulsion} Since the Hubbard $U$ is about 1 eV or
larger in strongly correlated materials, the electron-phonon
coupling must be large to stabilize on-site bipolarons and the
Holstein bipolaron mass appears very large, $m_{H}^{**}/m_0>1000$,
for realistic values of the phonon frequency.

 This estimate led some authors to the conclusion that the formation of
itinerant small polarons and bipolarons in real materials is
unlikely \cite{mel}, and high-temperature bipolaronic
superconductivity is impossible \cite{and2}. However, one should
note that the Holstein model is an extreme polaron model, and
typically yields the highest possible value of the (bi)polaron mass
in the strong coupling limit.  Many advanced materials with low
density of free carriers and poor mobility (at least in one
direction) are characterized by poor screening of high-frequency
optical phonons and are more appropriately described by the
long-range Fr\"ohlich electron-phonon interaction \cite{ale5}. For
this interaction the parameter $\gamma$ is less than 1
($\gamma\approx 0.3$ on the square lattice and $\gamma\approx 0.2$
on the triangular lattice), reflecting the fact that in a hopping
event the lattice deformation is partially pre-existent. Hence the
unscreened Fr\"ohlich electron-phonon interaction provides
relatively light small polarons, which are several orders of
magnitude lighter than small Holstein polarons.
\index{polaron!small!light}

As shown above  FCM is reduced to an extended Hubbard model
\index{Hubbard!model} with intersite attraction and suppressed
double-occupancy in the limit of high phonon frequency $\omega \geq
t $ and large on-site Coulomb repulsion. Then the Hamiltonian can be
projected onto the subspace of nearest neighbor intersite
bipolarons. In contrast with the crawler motion of on-site
bipolaron, the intersite bipolaron tunnelling
\index{bipolaron!tunnelling} is a crab-like, so that  its mass
scales linearly with the polaron mass ($m^{**}\approx 4m^*$ on the
staggered chain \cite{alekor2}) as confirmed numerically using CTQMC
algorithm by Kornilovitch \cite{kornil}. As a result , the crab
bipolarons could bose-condense already at the room temperature
\cite{jim}. \index{bipolaron!crab}

  We believe that the following recipe is worth investigating to
look for room-temperature superconductivity \cite{jim}:
\index{superconductivity!room temperature} (a) The parent compound
should be an ionic insulator with light ions to form high-frequency
optical phonons, (b) The structure should be quasi two-dimensional
to ensure poor screening of high-frequency c-axis polarized phonons,
(c) A triangular lattice is desirable in combination with strong,
on-site
 Coulomb repulsion to form the  superlight crab
bipolaron, and (d) Moderate carrier densities are required to keep
the system of small bipolarons close to the dilute regime. I believe
that most of these conditions are already met in cuprate
superconductors. As discussed above  there is  strong evidence for
3D bipolaronic BEC \index{bipolaron!Bose-Einstein condensate} in
cuprates from unusual upper critical fields and the
 electronic specific heat,  normal state pseudogaps  and anisotropy,
  normal state diamagnetism, the Hall-Lorenz numbers,  and the giant
proximity effect.

\section*{Acknowledgements}

I thank  A. F. Andreev,  J. P. Hague, V. V. Kabanov, P. E .
  Kornilovitch, and
J. H. Samson  for illuminating discussions and collaboration.  The
work was supported by EPSRC (UK) (grant no. EP/C518365/1).

%
%
%
%
%
%
%
%
%
%

%
%



\end{document}